\documentclass[twocolumn,trackchanges]{aastex63}

\usepackage{soul}
\usepackage{colortbl} 

\shorttitle{Expectations for SFHs from 2D-SPS of HST$+$JWST imaging}
\shortauthors{Garc\'{i}a-Argum\'{a}nez et al.}

\usepackage{ulem}
\usepackage{comment}
\usepackage{multirow}
\usepackage{subfigure}
\usepackage{enumerate}
\newcommand{\plotthree}[3]{
  \centering 
  \leavevmode  
  \includegraphics[trim=0 0 5 0,clip, scale=0.32]{#1}%
  \hfil
  \includegraphics[scale=0.32]{#2}%
  \hfil 
  \includegraphics[trim=0 0 84 0, clip, scale=0.32]{#3}
}%

\defcitealias{2017MNRAS.468..207S}{S+17}

\begin{document}

\title{Probing the earliest phases in the formation of massive galaxies\\with simulated HST$+$JWST imaging data from Illustris}

\correspondingauthor{\'{A}ngela Garc\'{i}a-Argum\'{a}nez}
\email{agarciaargumanez@ucm.es}

\author[0000-0002-8365-5525]{\'Angela Garc\'ia-Argum\'anez}
\affiliation{Departamento de F\'isica de la Tierra y Astrof\'isica, Facultad de CC F\'isicas, Universidad Complutense de Madrid, 28040, Madrid, Spain}
\affiliation{Instituto de F\'isica de Part\'iculas y del Cosmos IPARCOS, Fac$.$ CC F\'isicas, Universidad Complutense de Madrid, 28040 Madrid, Spain}

\author[0000-0003-4528-5639]{Pablo G. P\'{e}rez-Gonz\'{a}lez}
\affiliation{Centro de Astrobiolog\'ia (CSIC-INTA), Ctra de Ajalvir km 4, Torrej\'on de Ardoz, E-28850, Madrid, Spain}

\author[0000-0001-6150-2854]{Armando Gil de Paz}
\affiliation{Departamento de F\'isica de la Tierra y Astrof\'isica, Facultad de CC F\'isicas, Universidad Complutense de Madrid, 28040, Madrid, Spain}
\affiliation{Instituto de F\'isica de Part\'iculas y del Cosmos IPARCOS, Fac$.$ CC F\'isicas, Universidad Complutense de Madrid, 28040 Madrid, Spain}

\author[0000-0002-4226-304X]{Gregory F. Snyder}
\affiliation{Space Telescope Science Institute, 3700 San Martin Drive, Baltimore, MD 21218, USA}

\author[0000-0002-7959-8783]{Pablo Arrabal Haro}
\affiliation{NSF's National Optical-Infrared Astronomy Research Laboratory, 950 N. Cherry Ave., Tucson, AZ 85719, USA}

\author[0000-0002-9921-9218]{Micaela B. Bagley}
\affiliation{Department of Astronomy, The University of Texas at Austin, Austin, TX, USA}

\author[0000-0001-8519-1130]{Steven L. Finkelstein}
\affiliation{Department of Astronomy, The University of Texas at Austin, Austin, TX, USA}

\author[0000-0001-9187-3605]{Jeyhan S. Kartaltepe}
\affil{Laboratory for Multiwavelength Astrophysics, School of Physics and Astronomy, Rochester Institute of Technology, 84 Lomb Memorial Drive, Rochester, NY 14623, USA}

\author[0000-0002-6610-2048]{Anton Koekemoer}
\affiliation{Space Telescope Science Institute, 3700 San Martin Drive, Baltimore, MD 21218, USA}

\author[0000-0001-7503-8482]{Casey Papovich}
\affiliation{Department of Physics and Astronomy, Texas A\&M University, College Station, TX, 77843-4242 USA}
\affiliation{George P.\ and Cynthia Woods Mitchell Institute for Fundamental Physics and Astronomy, Texas A\&M University, College Station, TX, 77843-4242 USA}

\author[0000-0003-3382-5941]{Nor Pirzkal}
\affiliation{Space Telescope Science Institute, 3700 San Martin Drive, Baltimore, MD 21218, USA}

\author[0000-0001-7113-2738]{Harry C. Ferguson}
\affiliation{Space Telescope Science Institute, 3700 San Martin Drive, Baltimore, MD 21218, USA}

\author[0000-0003-3466-035X]{L. Y. Aaron\ Yung}
\affiliation{Astrophysics Science Division, NASA Goddard Space Flight Center, 8800 Greenbelt Rd, Greenbelt, MD 20771, USA}

\author[0000-0002-8053-8040]{Marianna Annunziatella}
\affiliation{Centro de Astrobiolog\'ia (CSIC-INTA), Ctra de Ajalvir km 4, Torrej\'on de Ardoz, E-28850, Madrid, Spain}

\author[0000-0001-7151-009X]{Nikko J. Cleri}
\affiliation{Department of Physics and Astronomy, Texas A\&M University, College Station, TX, 77843-4242 USA}
\affiliation{George P.\ and Cynthia Woods Mitchell Institute for Fundamental Physics and Astronomy, Texas A\&M University, College Station, TX, 77843-4242 USA}

\author[0000-0003-1371-6019]{M. C. Cooper}
\affiliation{Department of Physics \& Astronomy, University of California, Irvine, 4129 Reines Hall, Irvine, CA 92697, USA}

\author[0000-0001-6820-0015]{Luca Costantin}
\affiliation{Centro de Astrobiolog\'ia (CSIC-INTA), Ctra de Ajalvir km 4, Torrej\'on de Ardoz, E-28850, Madrid, Spain}

\author[0000-0002-4884-6756]{Benne W. Holwerda}
\affiliation{University of Louisville, Department of Physics and Astronomy, 102 Natural Science Building, Louisville, KY 40292, USA}

\author[0000-0001-8115-5845]{Rosa Mar\'ia M\'erida}
\affiliation{Centro de Astrobiolog\'ia (CSIC-INTA), Ctra de Ajalvir km 4, Torrej\'on de Ardoz, E-28850, Madrid, Spain}
\affiliation{Departamento de F\'isica Te\'orica, Universidad Aut\'onoma de Madrid, E-28049, Cantoblanco (Madrid), Spain}

\author[0000-0002-8018-3219]{Caitlin Rose}
\affil{Laboratory for Multiwavelength Astrophysics, School of Physics and Astronomy, Rochester Institute of Technology, 84 Lomb Memorial Drive, Rochester, NY 14623, USA}

\author[0000-0002-7831-8751]{Mauro Giavalisco}
\affiliation{Astronomy Department, University of Massachusetts, Amherst, MA 01003, U.S.A.}

\author[0000-0001-9440-8872]{Norman A. Grogin}
\affiliation{Space Telescope Science Institute, 3700 San Martin Drive, Baltimore, MD 21218, USA}

\author[0000-0002-8360-3880]{Dale D. Kocevski}
\affil{Department of Physics and Astronomy, Colby College, Waterville, ME 04901, USA}

\begin{abstract}

We use the Illustris-1 simulation to explore the capabilities of the \textit{Hubble} and \textit{James Webb Space Telescope} data to analyze the stellar populations in high-redshift galaxies, taking advantage of the combined depth, spatial resolution, and wavelength coverage. For that purpose, we use simulated broad-band ACS, WFC3 and NIRCam data and 2-dimensional stellar population synthesis (2D-SPS) to derive the integrated star formation history (SFH) of massive (M$_{\ast}>10^{10}\,$M$_{\odot}$) simulated galaxies at \mbox{$1<z<4$} that evolve into a local M$_{\ast}>10^{11}\,$M$_{\odot}$ galaxy. In particular, we explore the potential of HST and JWST datasets reaching a depth similar to those of the CANDELS and ongoing CEERS observations, respectively, and concentrate on determining the capabilities of this dataset for characterizing the first episodes in the SFH of local M$_{\ast}>10^{11}\,$M$_{\odot}$ galaxies by studying their progenitors at $z>1$. The 2D-SPS method presented in this paper has been calibrated to robustly recover the cosmic times when the first star formation episodes occurred in massive galaxies, i.e., the first stages in their integrated SFHs. In particular, we discuss the times when the first 1\% to 50\% of their total stellar mass formed in the simulation. We demonstrate that we can recover these ages with typical median systematic offset of less than 5\% and scatter around 20\%-30\%. According to our measurements on Illustris data, we are able to recover that local M$_{\ast}>10^{11}\,$M$_{\odot}$ galaxies would have started their formation by $z=16$, forming the first 5\% of their stellar mass present at $z \sim 1$ by $z=4.5$, 10\% by $z=3.7$, and 25\% by $z=2.7$.

\end{abstract}


\section{INTRODUCTION} \label{sec:intro}

The most massive galaxies in the nearby Universe are generally quiescent and present relatively old stellar populations (\citealt{2006ARA&A..44..141R} and references therein). This is consistent with the `downsizing' scenario, first introduced by \citet{1996AJ....112..839C}, and which states that the most massive galaxies started and completed their stellar mass assembly earlier than less massive ones. Nevertheless, the mechanisms involved and, in particular, the role of feedback in the downsizing scenario of galaxy formation remain unclear. We can summarize this downsizing scenario and its implications in our understanding of galaxy formation by the following key questions: (i) When did the first massive galaxies appear in the lifetime of the Universe? (ii) When did the first star formation stages happen in massive galaxy formation, and how fast and bursty were they? (iii) How did stellar mass assemble in massive galaxies? One way to address these questions is to look further and fainter in order to search for the most likely progenitors of local massive galaxies at different redshifts, and subsequently analyze their stellar populations, especially their Star Formation Histories (SFHs).

There are many theoretical studies that have addressed these issues related to the formation and assembly of massive galaxies by reconstructing their full galaxy assembly histories (e.g., \citealt{2007MNRAS.375....2D}). This kind of studies have provided strong evidence for a two-phase evolutionary scenario (e.g. \citealt{2010ApJ...725.2312O}) in which massive galaxies would have formed most of their stellar mass at high redshift ($z \lesssim 2$) via a first main dissipative phase of \textit{in-situ} star formation, followed by a secondary phase of multiple (dry) minor mergers that gradually increase the galaxy size. More recent studies based on both semi-analytical and hydrodynamical simulations have confirmed and polished this two-phase scenario by predicting the accreted stellar fractions of early-type galaxies and confirmed that the fraction of accreted material increases with lower redshift and higher stellar masses (e.g., \citealt{2013ApJ...766...38L, 2017ApJ...836..161L}; \citealt{2016MNRAS.463.3948D}; \citealt{2016MNRAS.458.2371R};  \citealt{2020MNRAS.497...81D}; \citealt{2021A&A...647A..95P}; \citealt{2022ApJ...935...37R}). 

From the observational point of view, the assembly history of a galaxy is imprinted on the properties of its stellar populations. Significant observational work has been conducted in the last few years to study the stellar populations of the most massive galaxies at $z>3$ from the integrated emission of photometrically-selected (and sometimes spectroscopically-confirmed) samples (e.g. \citealt{2017Natur.544...71G}, \citealt{2019ApJ...876..135A}, \citealt{2020ApJ...890L...1F, 2020ApJ...903...47F}, \citealt{2022ApJ...924...25M} and references therein). However, there are also spectroscopic and photometric studies that show that the physical properties of galaxies present systematic trends both radially, mainly, but also azimuthally in the local Universe (see, e.g., \citealt{2013A&A...553A.102D}, \citealt{2015Sci...348..314T}, \citealt{2016ApJ...828...27N}, \citealt{2017MNRAS.469.4063W}, \citealt{2018A&A...618A..64H}, \citealt{2020MNRAS.492.4149S}, \citealt{2022ApJ...926...81A}, \citealt{2022A&A...657A..95C}, \citealt{2022arXiv220303653B}). For these kinds of studies, the stellar population synthesis (SPS) modeling in two dimensions (2D) has become an essential tool to derive these subgalactic-scale resolved properties by analyzing their spatially-resolved emission. If applied to massive progenitors at high redshift, this spatially-resolved analysis can provide further clues on the role of different mechanisms (e.g., internal vs$.$ external, secular vs$.$ fast) on the evolution of massive galaxies, since the different mechanisms proposed act at different scales and have a different impact as a function of galactocentric distance (\citealt{2014MNRAS.438.1870D}, \citealt{2015MNRAS.450.2327Z}, \citealt{2015Sci...348..314T, 2016MNRAS.458..242T, 2018ApJ...859...56T, 2019MNRAS.487.5416T}, \citealt{2022arXiv220304979A}) and for different morphological components (e.g., \citealt{2021MNRAS.504.3058M}, \citealt{2022MNRAS.tmp.1686J} for local galaxies, and \citealt{2021ApJ...913..125C,2022ApJ...929..121C} for higher-redshift galaxies).

Still, we cannot ignore that the main drawback of the stellar population synthesis is the intrinsic degeneracies associated with the study of the emission from stars, even within individual pixels. Indeed, limitations in the spectral resolution, wavelength coverage and/or signal-to-noise of the data, result in strong degeneracies in the parameter space of physical properties such as age, star formation timescale, metallicity, or attenuation by dust. The correlations among these parameters are more or less difficult to disentangle depending on the value of each specific parameter (e.g., young ages, such as those expected in high-redshift galaxies, are less prone to some of these degeneracies), as well as the mentioned observational errors and wavelength coverage (see e.g$.$ \citealt{2002AJ....123.1864G}). 

The latter problems are especially difficult to tackle as we go to higher redshifts and/or study less massive galaxies, since galaxies are generally too faint to reach their continuum level with high spectral resolution. This makes the SPS analysis of progenitors of massive nearby galaxies at high redshift extremely challenging, as we cannot use a large enough sample of them reaching faint magnitudes to account for a significant fraction of their whole mass and, thus, obtain statistically robust results. These large samples are still harder to get when using spectroscopic (instead of photometric) data, which are resource-demanding and therefore relatively scarce. In addition, due the technological limitations of our observatories, it is hard to cover wide spectral ranges to account for different populations and effects such as dust extinction or other degeneracies.
 
In this regard, the \textit{James Webb Space Telescope} (JWST; \citealt{2006SSRv..123..485G}) will play an important part in overcoming these issues, thanks to its unprecedented sensitivity, high spatial resolution, wide wavelength coverage, sensitive to stars of different ages, and variety of spectral resolutions (from spectroscopic $R\sim3000$ to photometric $R\sim7$ and spectro-photometric $R=100$). Our approach is to use spatially-resolved multiwavelength broad-band data from JWST, combined with already existing broad-band data from the \textit{Hubble Space Telescope} (HST), to extract Spectral Energy Distributions (SEDs) of progenitors of nearby massive galaxies at high redshift and apply 2D-SPS modeling on them. The determination of stellar population parameters in two dimensions, and not only for the galaxy as a whole, will help us to analyze stellar mass distributions inside galaxies and to recover realistic integrated galactic SFHs. The power of analyzing the stellar populations in 2D resides in the fact that smaller regions of a given galaxy should have its observational properties driven by a more simple SFH, which can therefore be characterized with fewer parameters compared to that required to characterize an entire galaxy (which easily could count with several, quite different stellar populations). Analyzing the stellar content of high-redshift systems in 2D will be fundamental to identify the evolutionary stages a galaxy can undergo regarding its stellar content evolution and to determine when its assembly began.

This work aims at determining the robustness of spatially-resolved SFHs for massive galaxies at \mbox{$1 < z < 4$}, derived from the 2D-SPS analysis of simulated HST$+$JWST broad-band photometry, to infer when these massive progenitors began their stellar mass assembly. The reason for focusing on this $1 < z < 4$ redshift interval, for which JWST is expected to provide high-quality data, is that it includes the so-called ‘cosmic noon’, i.e., the epoch of the Universe where the cosmic star formation rate density history was maximum at $z \sim 2$ and where a considerable fraction of the local stellar mass was formed: half of the present-day stellar mass was formed before $z = 1.3$ (see \citealt{2014ARA&A..52..415M} and references therein). The results of this work will help us identify the first progenitors of massive galaxies around cosmic noon, an epoch of the Universe of great interest for many JWST studies in the upcoming years.

In order to establish a methodology and test its performance, we use galaxy images in different bands simulated by the Illustris Project. One of the most appealing features of recent cosmological simulations such as Illustris-1 (\citealt{2014Natur.509..177V}, \citealt{2015A&C....13...12N}) is that some of them provide synthetic stellar images generated for simulated galaxies at different redshifts in common broad-band filters from HST or JWST (\citealt{2015MNRAS.447.2753T}) and with ``realistic" morphologies (at least, relative to the morphologies provided by typical semi-analytical models). These images, when compared to the available information of the stellar particles for each galaxy in the simulation, make them the perfect benchmark to test how successful our 2D stellar population synthesis method is at recovering the early formation of massive galaxies at high redshift when upcoming JWST broad-band photometry data become available.

The present paper is structured as follows. In Section~\ref{sec:dataset}, we give a brief overview of the Illustris simulation and explain how the sample of simulated galaxies has been selected from the synthetic images. In Section~\ref{sec:data}, we present the processing of the images and the photometric method conducted to build SEDs in 2D. Section~\ref{sec:SPSmodel} describes the SPS modeling of the SEDs measured in 2D and the derivation of the integrated SFHs from these 2D-SPS analysis. In Section~\ref{sec:evaluation}, we evaluate the success of our method in recovering the earliest phases in the formation of each galaxy in the sample. Finally, in Sections~\ref{sec:results} and~\ref{sec:conclusions}, we present our results and outline our conclusions, respectively. 

Throughout this paper, we adopt the same cosmology as specified in the Illustris-1 simulation: $\Lambda$CDM cosmology with $\Omega_{m}$ = 0.2726, $\Omega_{\Lambda} = 0.7274$, $\Omega_{b} = 0.0456$, and $h = 0.704$. We assume a \cite{2001MNRAS.322..231K} initial mass function (IMF). All magnitudes presented in this work are calculated using the AB system (\citealt{1983ApJ...266..713O}).

\section{DATASET} \label{sec:dataset}

In this paper, we present a method to study the star formation history of massive, spatially resolved high-redshift galaxies using HST and JWST photometric data. We apply this method to a sample of galaxies simulated by the Illustris Project. Our aim is to assess the utility of HST plus JWST combined datasets for the analysis of the earliest evolutionary phases of nearby massive galaxies. For that purpose, our approach consists in selecting the progenitor galaxies at high redshift of nearby massive galaxies, and then analyzing their pixel-by-pixel and resulting integrated-galactic SFHs. In this Section, we describe the simulations, the sample, and the data.

\subsection{Illustris Simulation and synthetic deep-survey images}\label{subsec:illustris}

The Illustris Project is a set of hydrodynamical simulations of a (106.5 Mpc)$^{3}$ periodic cosmological volume that trace the evolution of dark matter, gas, stars, and supermassive black holes from $z=127$ to $z=0$ (\citealt{2014Natur.509..177V}, \citealt{2015A&C....13...12N}). Illustris comprises six different runs: Illustris-(1,2,3) and Illustris-(1,2,3)-Dark, where the former include a baryonic physical component in addition to the dark-matter content, and the latter the dark-matter component alone. We make use of the Illustris-1 run, the one with the highest-resolution box of these six runs in terms of the number of resolution elements and their masses: it follows the evolution of 1820$^3$ dark-matter particles with a mass of 6.26 $\times$ 10$^6$ M$_{\odot}$ each, and 1820$^3$ baryonic particles with an initial mass of 1.26 $\times$ $10^6$ M$_{\odot}$. The Illustris-1 volume contains $\sim 40,000$ galaxies at $z=0$ with more than 500 stellar particles $-$ roughly equivalent to galaxy stellar masses of M$_{\ast} > 5 \times 10^{8}$\,M$_{\odot}$ (\citealt{2015MNRAS.447.2753T}). Out of all these galaxies at $z=0$ in Illustris-1, 856 galaxies have stellar masses of M$_{\ast} > 10^{11}$\,M$_{\odot}$.

This work is based on the simulated galaxies and broad-band images from the Illustris-1 supplementary data catalogs published by \citet[hereafter \citetalias{2017MNRAS.468..207S}]{2017MNRAS.468..207S} and available via   MAST\footnote{MAST:\dataset[10.17909/T98385]{http://dx.doi.org/10.17909/T98385}}. These catalogs correspond to three synthetic deep survey square images in three different fields, which are labeled as field A, B, and C, each $2.8 \arcmin \times 2.8 \arcmin$ in size. As described in \citetalias{2017MNRAS.468..207S}, each of these three deep survey images has been created by applying the lightcone technique proposed in \cite{2007MNRAS.376....2K} to the periodic Illustris-1-simulation volume. This technique is based on replacing distant volume in the simulation with the output from an earlier cosmic time. It consists in first replicating the periodic cubic volume simulation until a desired comoving distance is reached and tracing a lightcone across all the simulation replications. Then, the output time of the simulation that fills the lightcone volume is varied as a function of the comoving distance, i.e., the distant volume in the lightcone is replaced with the output from an earlier cosmic time in the simulation. The three different survey images used in this work have been generated using the same lightcone geometry with three different orientations. Each of these lightcones contains unique galaxies in the simulation up to $z \sim 18$ with no repetition, although some of these galaxies can be repeated between the three different fields.

Finally, the physical information given by the output of the simulation at each redshift is processed with the spectral synthesis code SUNRISE  (\citealt{2006MNRAS.372....2J,2010MNRAS.403...17J}) to create synthetic images in arbitrary filters. This is done by assigning SEDs to the stellar particles according to their mass, age, and metallicity, and by projecting these quantities from the simulation space to pre-defined hypothetical cameras. For the creation of these images, Starburst99 stellar population models (\citealt{1999ApJS..123....3L}) were used, with a \cite{2001MNRAS.322..231K} initial mass function (IMF). The effect of dust absorption has been included in these mock images using a simple birth cloud plus diffuse dust model \citep{2000ApJ...539..718C}. This models the total effects of dust at our wavelengths of interest, without requiring expensive dust radiative transfer simulations for the entire fields.

These three synthetic deep-survey images, also called ``mock ultra deep fields'', are provided down to the spatial resolution of HST and JWST, among other observatories, in a wide range of broad-band filters, imitating the conditions of real galaxy surveys. Publicly available catalogs associated to these images include galaxies in the survey images whose rest-frame \textit{g}-band apparent magnitude is \textit{g} $< 30.0$ mag (19,347 galaxies).

\subsection{Sample Selection}\label{subsec:sample}

\begin{figure*}[t]
\epsscale{1.25}
\plotthree{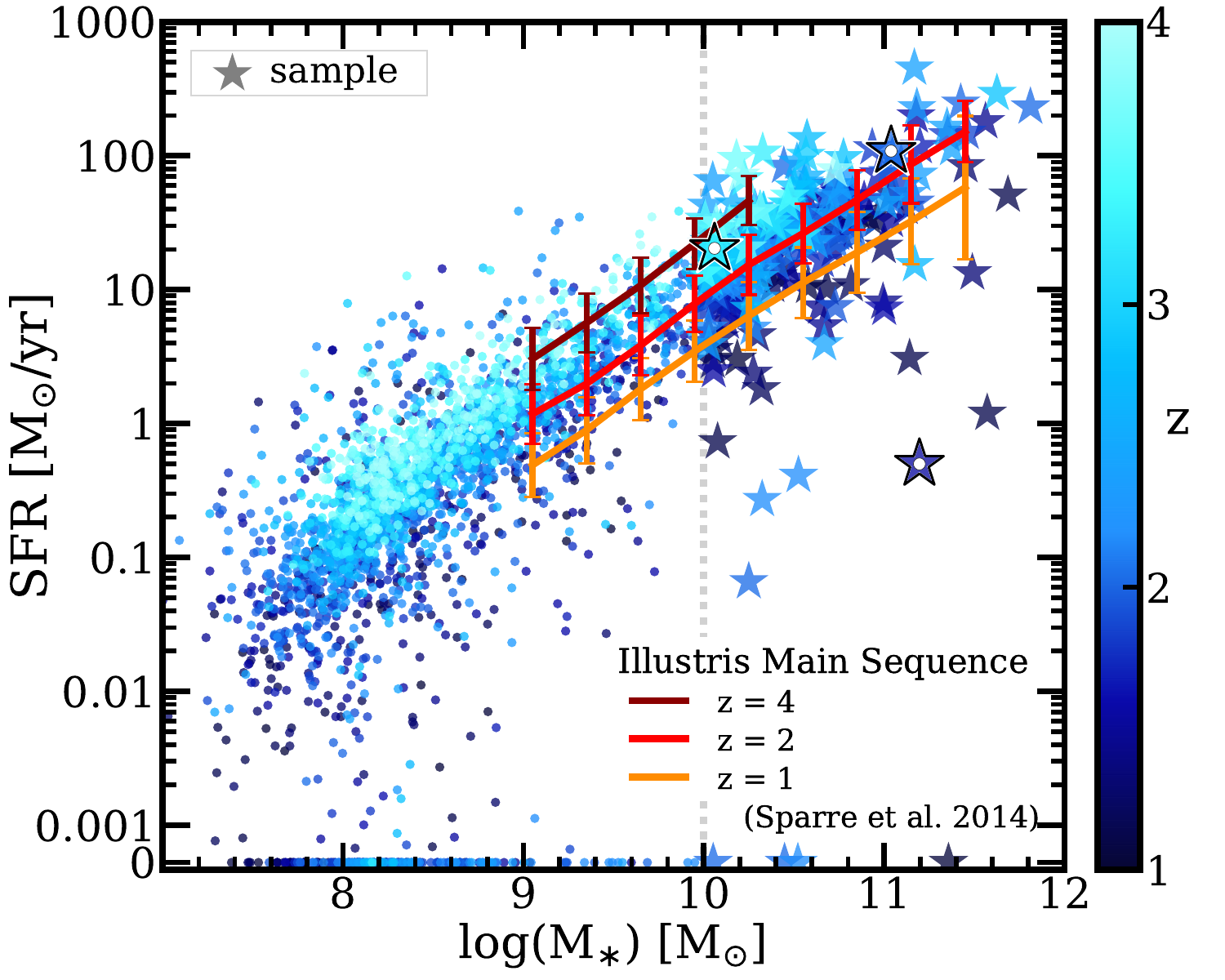}{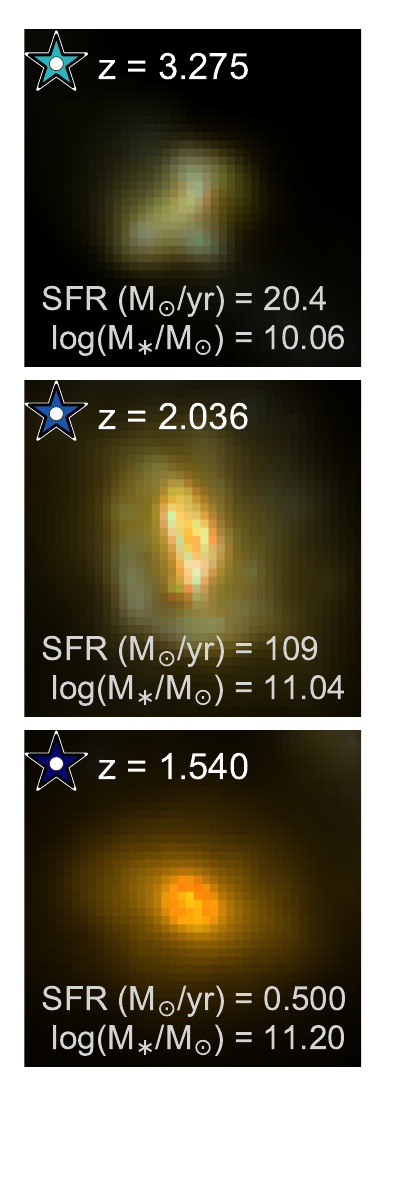}{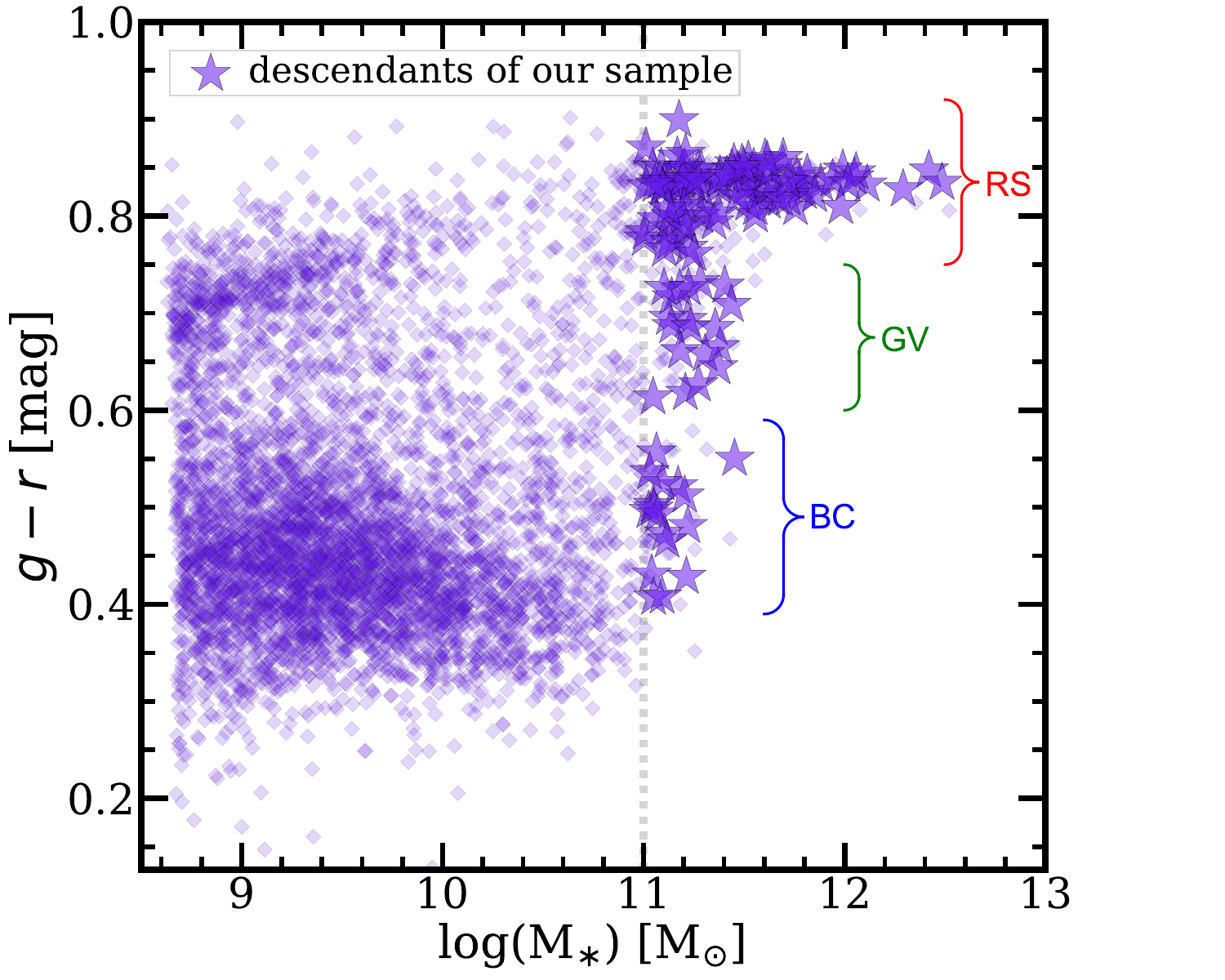}
\caption{Left panel: Main Sequence plot for those galaxies at $1<z<4$ included in the mock survey presented in \citetalias{2017MNRAS.468..207S} which are progenitors of very massive galaxies (M$_{\ast} > 10^{11}$ M$_{\odot}$) at $z=0$. All progenitors are color-coded by redshift. In this paper, we concentrate on the analysis of the most massive progenitors (M$_{\ast} > 10^{10}$ M$_{\odot}$), which are plotted with star symbols. The Main Sequence found for all the Illustris simulated galaxies at different redshifts (\citealt{2015MNRAS.447.3548S}) has also been plotted. Middle panel: postage stamp images for some representative examples of our galaxies (size $2.5\arcsec \times 2.5\arcsec$). These RGB images are created using ACS/F1814W, NIRCam/F200W, and NIRCam/F277W asinh-scaled images as B, G and R filters, respectively. The position of these galaxies in the left panel has been highlighted with a white dot inside. Right panel: $g-r$ color vs$.$ stellar mass diagram for descendants at $z=0$ of all galaxies at $1<z<4$ (independently of their stellar mass) in the \citetalias{2017MNRAS.468..207S} mock survey images. Descendants of the final sample of 221 galaxies analyzed in this paper are plotted as stars.
\label{fig:fig_sample}}
\end{figure*}

Starting from all the galaxies included in the catalogs of the \citetalias{2017MNRAS.468..207S} mock images, we select those at \mbox{$1 < z < 4$} which will evolve into a very massive (M$_{\ast} > 10^{11}$ M$_{\odot}$) descendant at $z=0$, as tracked forward in time via the Illustris merger trees \citep{2015MNRAS.449...49R}. This means that we select $1 < z < 4$ progenitors in the mock images of \mbox{$z=0$} very massive galaxies in the whole Illustris-1 simulation. Among all progenitor galaxies at \mbox{$1 < z < 4$} in the images, we restrict our analysis to massive galaxy progenitors with M$_{\ast} > 10^{10}$ M$_{\odot}$, for which very high quality JWST data will be available in the near future. This implies that, for the same 10$^{11}$ M$_{\odot}$ galaxy at $z=0$, all its massive progenitors at $1<z<4$ in the images are considered in our sample (e.g. two massive galaxies at $1<z<4$ could merge to form a M$_{\ast}>10^{11}$ M$_{\odot}$ galaxy at $z=0$). 

The left panel of Fig.~\ref{fig:fig_sample} shows the location in the star formation rate (SFR) vs$.$ stellar mass plane of all the progenitor galaxies at \mbox{$1<z<4$} in the \citetalias{2017MNRAS.468..207S} simulated images that have a M$_{\ast} > 10^{11}$ M$_{\odot}$ at $z=0$. These 2,994 progenitor galaxies shown in the left panel of Fig.~\ref{fig:fig_sample} represent 19\% of all galaxies located at $1<z<4$ in the images. This is equivalent to saying that 81\% of galaxies at $1<z<4$ in the images do not end up as a very massive galaxy at $z=0$. Out of that fraction of $1<z<4$ galaxies with a very massive descendant at $z=0$, our selected sample is composed of the 221 progenitor galaxies with M$_{\ast} > 10^{10}$ M$_{\odot}$ (represented by stars in Fig.~\ref{fig:fig_sample}, left panel). Among them, two of them appear simultaneously in two of the three mock ultra-deep fields with different orientations, and will be considered as independent galaxies in this work regarding the photometric analysis and the derivation of the SFHs. In fact, the initial selected sample from the images is composed of 248 galaxies, but only 221 (+ 2 repetitions) are kept until the final analysis. These 248 galaxies in the initial sample of massive progenitors represent the 64\% of all \mbox{M$_{\ast} > 10^{10}$ M$_{\odot}$} galaxies at $1<z<4$ in the images (388 galaxies), i.e., if we were to select all massive galaxies at $1<z<4$ in the images, only 64\% of them would actually become a M$_{\ast} > 10^{11}$ M$_{\odot}$ galaxy at $z=0$. Interestingly, this percentage rises to 68\% and 85\% when the mass cutoff is set to $10^{10.1}$ and $10^{10.5}$ M$_{\odot}$, respectively. In these cases, the number of galaxies in the sample would decrease from 388 galaxies to 305 for the $10^{10.1}$ M$_{\odot}$ cut and to 123 for $10^{10.5}$ M$_{\odot}$. In Fig.~\ref{fig:fig_sample}, we also show the galaxy Main Sequence for Illustris galaxies, as determined by \cite{2015MNRAS.447.3548S}.

The right panel of Fig.~\ref{fig:fig_sample} shows the $g-r$ color-stellar mass diagram for descendants at $z = 0$ of all galaxies at $1<z<4$ in the mock images of \citetalias{2017MNRAS.468..207S}. Only descendants with log(M$_{\ast}$/M$_{\odot}$)$\gtrsim 8.68$ are shown, for which integrated photometry data is available (5,498 galaxies). Since descendants at $z=0$ of galaxies in \citetalias{2017MNRAS.468..207S} mock images have been traced via the Illustris merger tree and do not appear in these survey images, the integrated photometry magnitudes used to build this diagram have been extracted from the \cite{2015MNRAS.447.2753T} synthetic individual galaxy images. But these mock galaxy images are only available for $z=0$ descendants with a minimum stellar mass of log(M$_{\ast}$/M$_{\odot}$)$\gtrsim 8.68$ ($\sim 63\%$ of all the 8,768 $z=0$ descendants of $1<z<4$ galaxies in the images).

Our final 221 galaxies selected from these images evolve to 132 (unique) very massive galaxies at $z=0$, which means that some galaxies in our final sample at $1 < z < 4$ have the same descendant at $z=0$. These descendants are called $z=0$ descendants of our main sample, hereafter, and are shown as violet stars in the right panel of Fig.~\ref{fig:fig_sample}. We note that 51\% of all $z=0$ M$_{\ast} > 10^{11}$ M$_{\odot}$ with a progenitor galaxy at $1<z<4$ in the images do not come from any of the galaxies in our initial sample of 248 $1<z<4$, M$_{\ast} > 10^{10}$ M$_{\odot}$ systems. The main progenitor at $z>1$ in the images of those nearby massive galaxies presents a typical stellar mass $\log ($M$_{\ast}$/M$_{\odot}) = 9.12^{9.65}_{8.48}$ and a typical redshift of $z=2.71^{3.42}_{2.01}$ (median values and first and third quartiles).

The 221 galaxies in the sample account for 28\% of the total stellar mass present in their 132 $z=0$ descendants, where the 22 galaxies at $3<z<4$ account for 8.1\% of the stellar mass of their descendants, the 94 galaxies at $2 < z < 3$ for 23\%, and the 105 galaxies at \mbox{$1<z<2$} for 33\%. These descendants are predominantly Red Sequence galaxies (RS; approximately 73\%), but there are also a few galaxies in the Green Valley (GV; 14\%) and in the Blue Cloud (BC; 13\%). In the case of the M$_{\ast} > 10^{11}$~M$_{\odot}$ descendants of progenitor galaxies at $1<z<4$ which are not massive, their location in the diagram is similar, but with a higher fraction of them in the BC: 63\% in the RS, 16\% in the GV, and 21\% in the BC. 

\begin{figure}[t]
\epsscale{1.15}
\plotone{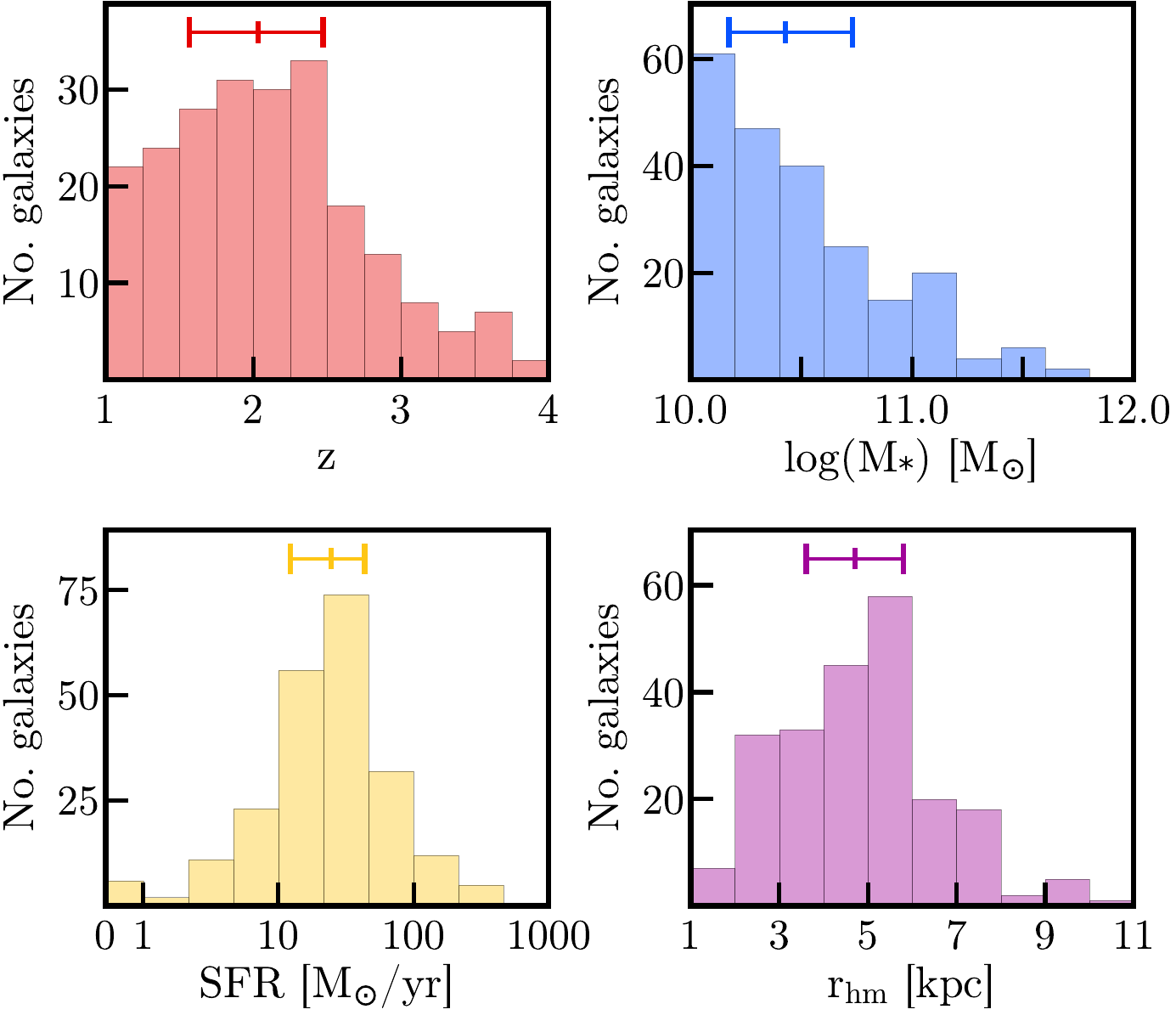}
\caption{Histograms for our $1<z<4$ sample of 221 simulated galaxies: redshift (top left panel), total stellar mass (top right), SFR (bottom left), and stellar half-mass radius (bottom right). These properties have been extracted from the Illustris-1 database. Median and quartiles are shown as segments on the top. 
\label{fig:fig_sample_hists}}
\end{figure}

The galaxies in our sample are projected into three $2.8 \times 2.8$ arcmin$^2$ area \citepalias{2017MNRAS.468..207S}. The typical stellar mass, star-formation rate (SFR), redshift and stellar half-mass radius (i.e$.$, the radius enclosing half of the total stellar mass of the galaxy; r$_{\text{hm}}$) of galaxies in the sample are \mbox{$\langle \log($M$_{\ast}) \rangle = 10.4^{10.7}_{10.1}$ M$_{\odot}$}, $\langle $SFR$ \rangle = 25^{43}_{12}$ M$_{\odot}$/yr, $ \langle z \rangle =2.0^{2.5}_{1.6}$, and $ \langle $r$_{\text{hm}} \rangle = 4.7^{5.8}_{3.6}$ kpc (median and quartiles), respectively. Fig.~\ref{fig:fig_sample_hists} shows the histograms of these properties for our sample of 221 galaxies.

\newpage
\section{PHOTOMETRIC DATA FROM THE ILLUSTRIS SIMULATION}\label{sec:data}

In this section, we describe how the simulated deep survey images have been processed to obtain SEDs for all galaxies in our sample. We measure integrated-light photometry for each galaxy, and we also consider spatially resolved SEDs, all constructed with data in several HST and JWST filters. Apart from simulated images, we also use redshifts and physical properties extracted from the Illustris database. Throughout this paper, we refer to $`$ground-truth' or $`$reference' properties as those galaxy properties derived either from the catalogs associated to the \citetalias{2017MNRAS.468..207S} mock images or from the Illustris database, in contrast to the properties obtained from SED fits to stellar population synthesis models for different regions in each galaxy, what we call 2D-SPS-derived properties.

\subsection{Photometric broad-band filters}

As described in Section~\ref{subsec:illustris}, this work uses the Illustris-1 “mock ultra-deep fields” from \citetalias{2017MNRAS.468..207S}, each 2.8 arcmin in size. These synthetic deep survey images are available to be used in 34 broad-band filters onboard the HST/ACS and WFC3, JWST/NIRCam and MIRI, and Roman/WFI. For this work, we consider 15 HST and JWST broad-band filters in the optical and near-infrared (see Table~\ref{tab:filters}), which will be available for several cosmological fields covered by HST legacy projects such as the Cosmic Assembly Near-infrared Deep Extragalactic Legacy Survey fields (CANDELS; \citealt{2011ApJS..197...35G}; \citealt{2011ApJS..197...36K}), and JWST Guaranteed Time Observations (GTO), Cycle 1 Guest Observers (GO1) or Early Release Science (ERS) programs such as the MIRI Deep Survey (\citealt{2017jwst.prop.1283N}), the JWST Advanced Deep Extragalactic Survey (JADES; \citealt{2018ApJS..236...33W}, \citealt{2019BAAS...51c..45R}), the Public Release IMaging for Extragalactic Research (PRIMER; \citealt{2021jwst.prop.1837D}) or the Cosmic Evolution Early Release Science (CEERS; \citealt{2017jwst.prop.1345F}). The filterset covers the optical through mid-infrared observed spectral region, corresponding to rest-frame UV to near-infrared wavelengths at $1<z<4$, an adequate range to study the emission from young and old stellar populations (i.e., providing information about the SFH). The survey images released by \citetalias{2017MNRAS.468..207S} are noise-free synthetic images with different spatial resolution depending on the instrument and telescope used (see Table~\ref{tab:filters}) and available with or without considering the point spread function (PSF) degradation for each filter. For this work, we use as starting images those without the PSF model applied, with pixel scales provided in Table~\ref{tab:filters}. 

\begin{deluxetable}{cccrcc}[ht]
\tablenum{1}
\tabletypesize{\footnotesize}
\setlength{\tabcolsep}{3.5pt}
\tablecaption{Photometric broad-band images used in this work\label{tab:filters}}
\tablehead{ \colhead{} & \colhead{} & \colhead{$\mathrm{\lambda}\mathrm{_{central}}$} & \colhead{Width} & \colhead{Pixel scale\tablenotemark{\scriptsize a}} & \colhead{ 5$\mathrm{\sigma}$ depth\tablenotemark{\scriptsize b} } \\[-0.43cm]
\colhead{Instrument} & \colhead{Filter} & \colhead{} & \colhead{} & \colhead{} & \colhead{}\\[-0.43cm] 
\colhead{} & \colhead{} & \colhead{($\mathrm{\mu}$m)} & \colhead{($\mathrm{\mu}$m)} & \colhead{(\arcsec/pix)} & \colhead{(mag)}}
\decimals
\startdata
ACS & F435W & 0.4318 & 0.0993 & 0.03 & 27.3 \\
@\textit{HST} & F606W & 0.5919 & 0.2225 & 0.03 & 27.4 \\
 & F775W & 0.7693 & 0.1491 & 0.03 & 26.9 \\
 & F814W & 0.8057 & 0.2358 & 0.03 & 27.2 \\
 & F850LP & 0.9036 & 0.2092 & 0.03 & 26.5 \\
\hline
\phantom{s}WFC3 & F105W & 1.0585 & 0.2653 & 0.06 & 26.1\tablenotemark{\scriptsize c} \\
@\textit{HST} & F125W & 1.2471 & 0.2867 & 0.06 & 26.1 \\
 & F140W & 1.3924 & 0.3760 & 0.06 & 25.6 \\
 & F160W & 1.5396 & 0.2744 & 0.06 & 26.4 \\
\hline
NIRCam & F115W & 1.1512 & 0.2426 & 0.032 & 29.2\\
@\textit{JWST} & F150W & 1.5017 & 0.3309 & 0.032 & 28.9\\
 & F200W & 1.9905 & 0.4654 & 0.032 & 29.0\\
 & F277W & 2.7861 & 0.7117 & 0.065 & 29.2\\
 & F356W & 3.5594 & 0.8163 & 0.065 & 29.0\\\
 & F444W & 4.4457 & 1.1197 & 0.065 & 28.6 \\
\enddata
\vspace{0.2cm}\scriptsize{\textbf{Notes:}}
\vspace{-0.1cm}\tablenotetext{a}{Pixel size in original images. Final (matched) pixel size is 0.06$\arcsec$.}
\vspace{-0.2cm}\tablenotetext{b}{HST depths from the CANDELS/3D-HST catalogs (\citealt{2014ApJS..214...24S}): median $5\sigma$ depth calculated from the errors of objects in the final catalogs (apertures of 0.7\arcsec) for all 5 CANDELS fields. JWST depths corresponding to the CEERS proposal: values represent the planned 5$\sigma$ point source depths per filter, assuming a total integration time of $2867\,$s for all NIRCam filters except for F115W ($5734\,$s).}
\vspace{-0.2cm}\tablenotetext{c}{This limiting magnitude is not included in the CANDELS/3D-HST catalogs from \citealt{2014ApJS..214...24S}. We assume the same value as F125W, since the magnitudes for this band are slightly fainter/brighter than those of F105W in the CANDELS survey.}
\vspace{-0.75cm}
\end{deluxetable}

\subsection{Photometric Measurements}\label{subsec:phot}

We process the original idealized (without PSF and noise) images for HST and JWST filters in order to imitate CANDELS data (\citealt{2011ApJS..197...35G}; \citealt{2011ApJS..197...36K}) and the recently-begun CEERS observations (\citealt{2017jwst.prop.1345F}), respectively. The reason for choosing the CEERS survey as a test case is that, although depths for subsequent JWST surveys are expected to be better, CEERS is expected to provide one of the first publicly available datasets for deep-field JWST observations. Cycle 1 GO / Treasury programs such as PRIMER are expected to achieve similar depths.

First, we match the pixel size of the images to that of HST/WFC3 (0.06 arcsec). Second, we convolve the image for each filter  with the PSF model for the WFC3/F160W filter (FWHM $\thickapprox$ 0.19 arcsec). Finally, we add Gaussian sky noise to the HST and JWST images to achieve the same depth as in the CANDELS and CEERS observations, respectively. We degrade HST images adding noise in order to match the median background rms noise measured around CANDELS massive galaxies (M$_{\ast} > 10^{10}$ M$_{\odot}$) at $1 < z < 4$. For this calculation, we use actual regions covered by CANDELS/Wide, CANDELS/Deep (\citealt{2011ApJS..197...35G}, \citealt{2011ApJS..197...36K}), and the Hubble Ultra Deep field (HUDF; \citealt{2006AJ....132.1729B}, \citealt{2010ApJ...709L..16O}, \citealt{2013ApJ...763L...7E}, \citealt{2013ApJS..209....3K}, \citealt{2013ApJS..209....6I}). For JWST images, the sky noise is estimated using the official JWST exposure time calculator \texttt{Pandeia}\footnote{\url{https://jwst.etc.stsci.edu/}. Pandeia Version: 1.7} (\citealt{2016SPIE.9910E..16P}), by measuring the predicted signal-to-noise ratio (SNR) as a function of the pixels surface-brightness in CEERS observations for the selected JWST broad-band filters. The  initial spatial resolution of the images for each filter (before registering) and final depths (after the noise addition) are given in Table~\ref{tab:filters}. The limiting magnitude has been estimated (without aperture correction) from the sky rms value adopting a circular aperture with a fixed radius of 0.2$\arcsec$ at a 5$\mathrm{\sigma}$ level. 

We use these processed (registered, PSF-matched, and sky noise-added) deep survey images to measure photometry. To do that, we first generate a segmentation map by running \texttt{SExtractor} (\citealt{1996A&AS..117..393B}) on the WFC3/F160W image using the following parameters: we filter images with a gaussian kernel FWHM 3 pixels, set the minimum area for detections to 20 (a bit smaller than the PSF FWHM), the number of deblending sub-thresholds to the maximum (64), the contrast to $5\times 10^{-4}$, and local background to a size of at least 4 times the largest galaxy in our sample. The results did not vary much when changing the relevant parameters given that we are dealing with relatively bright galaxies. In general, there is good agreement between sources detected in the segmentation map and galaxy positions given by the Illustris catalogs based on the original images. Then, this segmentation map is combined with the information of the sample in the Illustris catalogs to create circular galaxy apertures that enclose the integrated emission from all galaxies in the sample (see~\ref{subsubsec:iphot} below). We choose to use galaxy positions given by the Illustris catalogs instead of the ones provided by \texttt{SExtractor} in order to make the future comparison between our results derived from 2D-photometry and from the simulated particles in the galaxy more fair (otherwise, the centers for the galaxy photometric apertures would be shifted from the galaxy centers given by the simulated particles). 
 
We build galaxy SEDs by measuring multi-wavelength photometry in two different ways: inside these circular galaxy apertures, which enclose the entire flux of the galaxy (hereinafter ``integrated photometry"), and for small parts of the galaxy defined after creating a grid inside this circular aperture with cell size equal to 3$\times$3 pixels, roughly the area of the FWHM of the PSF-homogenized dataset (``2D photometry", hereafter). In both cases, the flux uncertainty is estimated as $\sqrt{N}\times \sigma$, where $N$ is the number of pixels within the aperture and $\sigma$ the rms of the sky for that filter.

\subsubsection{Integrated Photometry}\label{subsubsec:iphot}

In order to define the integrated-photometry apertures, we use both the segmentation map and information extracted from the Illustris database, in particular, the galaxy centers and stellar half-mass radii. The apertures are first defined in the segmentation map around galaxy centers with an initial radius equal to twice the stellar half-mass radius ($r_{\text{hm}}$). We decided to use the $r_{\text{hm}}$ values from the Illustris database because many of the galaxy properties provided by these simulations refer to this radius (and to $2 \times r_{\text{hm}}$) and, typically, papers compare the results based on observations to what Illustris provides for $2\times r_{\text{hm}}$ (see, e.g$.$, \citealt{2014MNRAS.444.1518V}, \citealt{2015MNRAS.447L...6S}, \citealt{2016ApJ...833..158C}, \citealt{2018MNRAS.479.4004E}, \citealt{2020ApJ...889...93V}) . With the following procedure, we test how those apertures compare to what is directly measured in the simulated images. The initial radius is first reduced to minimize the contamination from other sources until 80\% of the pixels within the aperture belong to the considered galaxy, i.e., we allow up to 20\% of pixels from neighbors. If the number of sky pixels (defined as those not belonging to any source) within this new aperture is greater than 10\%, we further reduce the radius to decrease that number below 10\%. We refer to the final photometric aperture radius as $r_{\text{phot}}$ hereafter. We discard galaxies (from the initial selected sample of 248 galaxies) whose final aperture radius is $r < 0.75\times r_{\text{hm}}$. This showed to be effective in removing sources overdeblended by \texttt{SExtractor} and galaxies whose initial aperture (with $r=2\times r_{\text{hm}}$) is partially beyond the edges of the simulated images. We also discard galaxies which present surface brightness values fainter than 25 mag/arcsec$^{2}$ in WFC3/F160W in their brightest pixel within the final aperture.

Typically, the fraction of pixels within the final integrated photometric apertures belonging to other nearby galaxies is very low, less than 1\% for 75\% of the sample of 221 galaxies. The maximum percentage of pixels from neighbor galaxies within the integrated apertures is $\sim$20\%, but only 27 out of 221 galaxies have more than 10\% of pixels belonging to other sources.

\begin{figure}[t]
\includegraphics[width=\columnwidth, trim=2 0 4 0,clip]{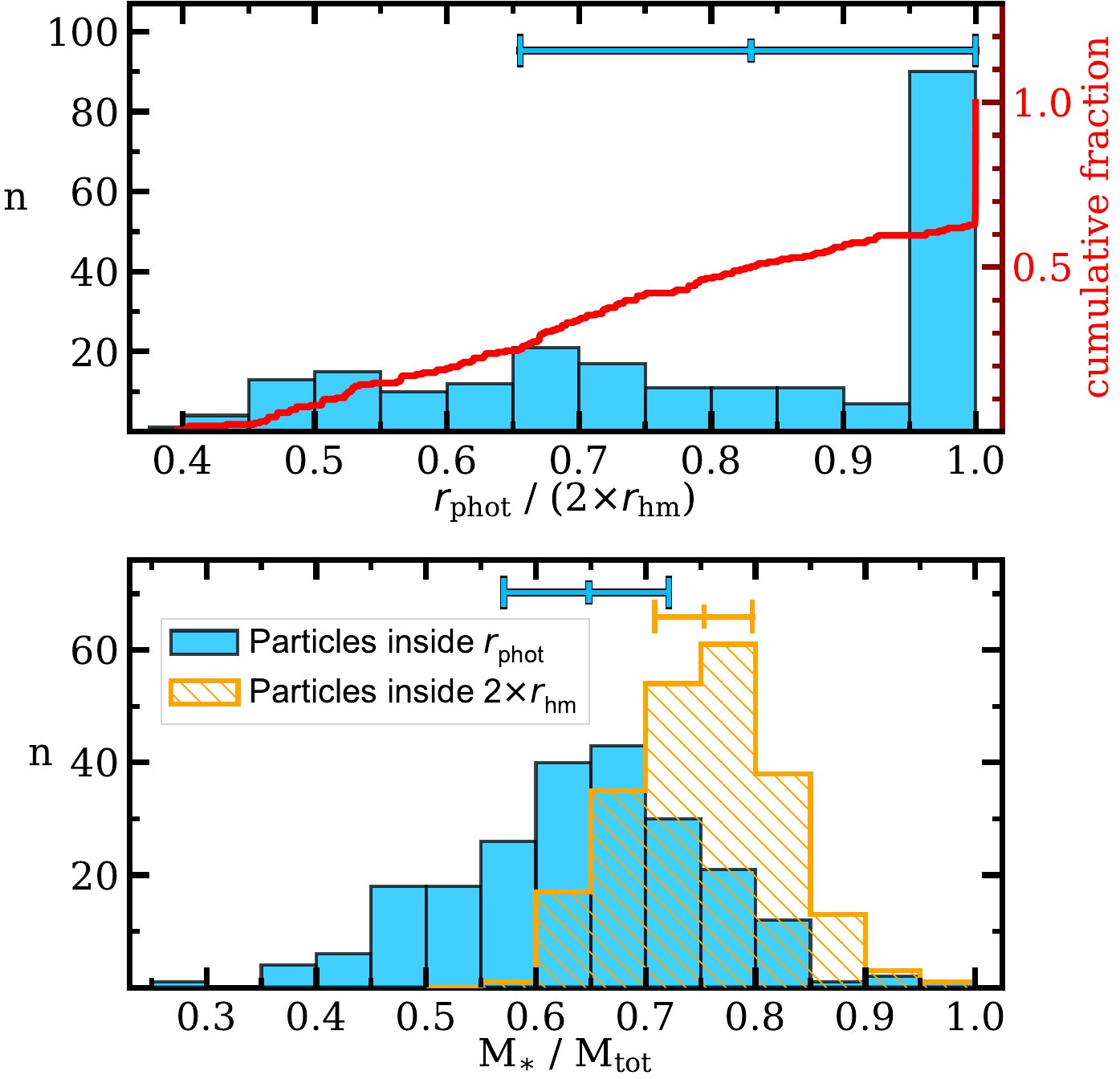}
\caption{Top panel: histogram for the radii of the photometric apertures of the 221 galaxies in the final sample as a fraction of $2\times r_{\text{hm}}$ radius, the typical radius used by Illustris to compare with observations. The cumulative fraction of galaxies in shown as a red line. Bottom panel: histogram for the fraction of stellar mass enclosed by the photometric apertures (blue filled histogram) and for a radius of $2\times r_{\text{hm}}$ (orange hatched) with respect to the total stellar mass in the galaxy. In both cases, we neglect neighboring galaxies. These total masses have been extracted from the simulated particles belonging to each galaxy in the Illustris database. At the top of both panels, we show the median and quartiles for each histogram.
\label{fig:rphot_vs_r2hm}}\vspace{-0.3cm}
\end{figure}

In Fig.~\ref{fig:rphot_vs_r2hm} (top panel) we show the histogram for the final radii of the apertures as a fraction of $2\times r_{\text{hm}}$ for the 221 galaxies in the final sample. This means we do not have apertures in this histogram at $r/(2\times r_{\text{hm}})<0.375$, due to the threshold imposed on the final apertures to be kept: $r > 0.75\times r_{\text{hm}}$. We find the median and quartiles for $r_{\text{phot}}/(2\times r_{\text{hm}})$ are $0.83^{1.00}_{0.66}$. It can be noticed that a considerable number of galaxies presents radii very similar or equal to the (starting) aperture radius of $2\times r_{\text{hm}}$: \mbox{$\sim 40$\%} and $\sim 37$\% of galaxies have $r_{\text{phot}}/(2\times r_{\text{hm}})\geq0.95$ and 0.999, respectively. On the contrary, only $\sim 8$\% of galaxies have aperture radii smaller than $r_{\text{hm}}$. But assuming $2\times r_{\text{hm}}$ as the best aperture to compare our results with simulations has demonstrated not to work for a significant number of galaxies. 

The stellar masses enclosed by our final apertures, a fraction of the total stellar mass of each galaxy, can be seen in the bottom panel of Fig.~\ref{fig:rphot_vs_r2hm}. As a comparison, we also show the histogram for the fraction of stellar mass enclosed by a radius of $2\times r_{\text{hm}}$. For both radii, the masses have been extracted from the Illustris database, i.e., they are ground-truth masses calculated by adding up the simulated stellar particles belonging to each galaxy which are closer to the galaxy center than the radii considered. The typical percentages of the stellar masses enclosed by these two apertures are (median and quartiles) $65_{57}^{72}$\% for the final photometric apertures and $75_{71}^{80}$\% for a radius of $2\times r_{\text{hm}}$. These percentages refer to total stellar masses provided by the Illustris database or using all particles (in both cases, referring to the whole dark matter halo, i.e., they include regions with very low mass surface densities, whose emission is well below our observational limits). This procedure aims at reproducing that to be applied to actual HST+JWST observations, so we should be able to evaluate its impact on reproducing ground-truth properties.

To measure the integrated flux for each galaxy, we only consider pixels whose center lies within the final circular aperture and which do not belong to other sources (according to the segmentation map). The fluxes of these pixels within the aperture (considered galaxy and sky pixels) are added together to build the integrated SED. In Table~\ref{tab:imag} we show the typical integrated magnitudes (median and quartiles) of our 221 galaxies in several filters and in different redshift bins.

\begin{deluxetable}{ccccc}[h]
\tablenum{2}
\setlength{\tabcolsep}{3.9pt}
\tablecaption{Integrated magnitudes of our $1<z<4$ sample \label{tab:imag}}
\tablehead{ \colhead{} & \colhead{} & \multicolumn3c{$m_{\mathrm{AB}}$ (mag)} \\[-0.43cm]
\colhead{Instrument} & \colhead{Filter} & \colhead{} & \colhead{} & \colhead{}\\[-0.43cm] 
\colhead{} & \colhead{} & \colhead{$1<z<2$} &  \colhead{$2<z<3$} & \colhead{$3<z<4$}}
\decimals
\startdata
HST/ACS\phantom{sS} & F435W & $23.0^{23.6}_{22.1}$ & $23.5^{24.6}_{22.9}$ & $25.5^{26.4}_{24.9}$ \\
HST/WFC3 & F160W & $21.7^{22.4}_{21.0}$ & $22.5^{23.1}_{21.9}$ & $23.3^{24.1}_{23.2}$ \\
\hline
JWST/NIRCam & F115W & $22.0^{22.6}_{21.1}$ & $23.1^{23.9}_{22.5}$ & $23.8^{24.3}_{23.3}$ \\
& F200W & $21.7^{22.4}_{20.9}$ & $22.5^{23.1}_{21.9}$ & $23.0^{23.7}_{22.8}$ \\
& F356W & $21.4^{22.2}_{20.7}$ & $22.3^{22.8}_{21.7}$ & $23.0^{23.7}_{22.8}$ \\
\enddata 
\vspace{0.2cm}
\scriptsize{\textbf{Note:}} Median values, first and third quartiles of the integrated magnitude measured in different redshift bins and for several bands.
\end{deluxetable}

\vspace{-0.8cm}

\subsubsection{2D Photometry}

Separately, we also define a grid inside the circular aperture with cell size equal to 3$\times$3 pixels ($0.18''\times 0.18''$) and we measure SEDs for each of these spatial resolution elements. We only consider cells inside the aperture which have at least one pixel that belongs to the galaxy we are measuring. If there are any pixels in the cells belonging to other galaxies (according to the segmentation map), the value of these pixels is replaced by sky pixels by adopting the same Gaussian sky noise distribution previously added to the images. Additionally, we only keep SEDs for our analysis from cells with a SNR $>$ 3 in at least 5 bands and with surface brightness brighter than 25 mag/arcsec$^{2}$ in WFC3/F160W. Cells that do not satisfy these conditions are discarded. The reason for imposing this surface brightness limit of 25 mag/arcsec$^{2}$ is to deal with the problem of the larger-than-observed galaxy sizes in Illustris galaxies with M$_{\ast} \lesssim 10^{10.7}$ M$_{\odot}$, which present larger half-light radii and more extended discs than real observed galaxies (\citealt{2015MNRAS.454.1886S}).

\begin{figure}[t]
\centering
\includegraphics[width=0.96\columnwidth, trim=0 -3.5 0 -8,clip]{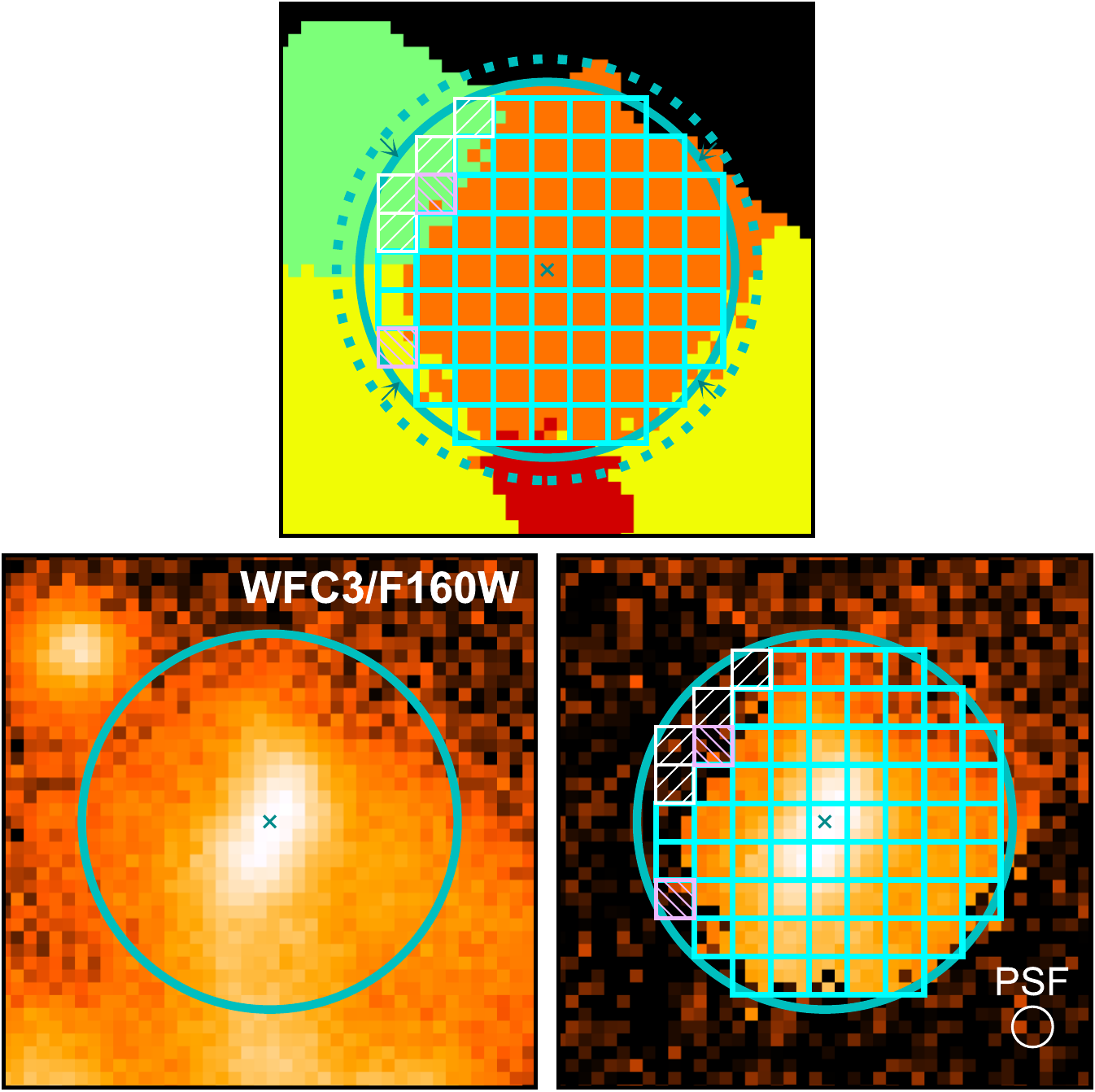}
\caption{Postage stamp images (size $2.5\arcsec \times 2.5\arcsec$) for one of our galaxies: Illustris-1\_066\_0000006 at z = 2.192 (Field A of \citetalias{2017MNRAS.468..207S} images). This galaxy code stands for galaxy id 6 and snapshot 66 ($z=2.21$) in the Illustris-1 simulation. Top panel: Segmentation map, with colored regions delimiting the galaxy (in orange), three nearby galaxies (green, yellow, and red), and sky pixels (black). We also show the integrated photometric aperture (dark cyan solid line) and the starting aperture of $r=2 \times r_{\text{hm}}$ used to calculate the former (dotted line). Inside the integrated aperture, we show the grid where the 2D photometry is measured: the cells that are finally kept for the 2D-SPS analysis are shown in cyan, cells discarded for not fulfilling the SNR and surface brightness criterion as pink hatched squares, and cells discarded for not including any pixel from the considered galaxy as white hatched squares. Bottom left panel: WFC3/F160W image (already registered, PSF-matched, and sky noise-added) with the integrated aperture. Bottom right: WFC3/F160W image where the values of the pixels from nearby galaxies have been masked by replacing them with the same Gaussian sky noise distribution previously added to the images. We also show the integrated aperture and the grid.}
\label{fig:apertures}
\end{figure}

Fig.~\ref{fig:apertures} shows an example of the integrated aperture and the grid for one of the galaxies in the sample. We also include in this figure the segmentation map around this galaxy and the initial integrated aperture of $r=2 \times r_{\text{hm}}$ (dotted line). We show the WFC3/F160W image for this galaxy before and after replacing the values of the pixels of nearby galaxies by random values drawn from the same Gaussian sky noise distribution previously used to add the noise in this image. The initial grid covers all the region within the final integrated aperture, but we only keep cells in cyan for the stellar population analysis, as described in the previous paragraph.

The typical number of cells per galaxy decreases as we move to higher redshifts, as expected given the smaller size of high-redshift galaxies (e.g., \citealt{2004ApJ...611L...1B}, \citealt{2010ApJ...709L..21O}, \citealt{2013ApJ...777..155O}). Our galaxies present (median and quartiles) 98$^{136}_{\phantom{1}69}$ cells at $1< z < 2$, 52$^{86}_{37}$ at $2 < z < 3$, and 24$^{29}_{21}$ at $3 < z < 4$. We also calculate the median SNR of the galaxy cells included in the analysis in different redshift bins, by calculating the SNR as the ratio between the measured flux and its uncertainty in these cells. Median and quartile values for ACS/F435W (WFC3/F160W) are 31$^{48}_{22}$ (30$^{52}_{22}$), at $1< z < 2$, 30$^{40}_{21}$ (27$^{38}_{22}$) at $2 < z < 3$, and 17$^{23}_{10}$ (24$^{31}_{21}$) at $3 < z < 4$. For F115W, F200W and F356W NIRCam bands, the median SNR of the galaxy cells are: 62$^{93}_{51}$, 72$^{110}_{\phantom{1}61}$ and 92$^{132}_{\phantom{1}77}$, respectively, at $1< z < 2$, 43$^{54}_{34}$, 66$^{88}_{60}$ and 88$^{110}_{\phantom{1}77}$ at $2 < z < 3$, and 53$^{69}_{48}$, 81$^{91}_{74}$ and 91$^{108}_{\phantom{1}80}$ at $3 < z < 4$.

\subsection{Redshifts}

Galaxy redshifts are taken from the Illustris database.  Specifically, we use the $`$inferred redshift' values in the catalogs that were used to generate the simulated images in \citetalias{2017MNRAS.468..207S}. This redshift value corresponds to the inferred cosmological redshift obtained considering the contribution of both the true cosmological redshift and the galaxy peculiar velocity. A discussion about the possible differences in the derived stellar population properties due to uncertainties in the redshifts of the corresponding observed galaxies is beyond the scope of this paper. Although photo-z uncertainties can be substantial, it will be likely that future spectroscopic redshift samples from both JWST and the \textit{Atacama Large Millimeter/submillimeter Array} (ALMA) will increase the fraction of galaxies with spectroscopic redshifts and, at the same time, improve the photo-z accuracy at such high redshifts.

\begin{figure}[t]
\centering
\includegraphics[width=\columnwidth, trim=0 7 54 63.5, clip]{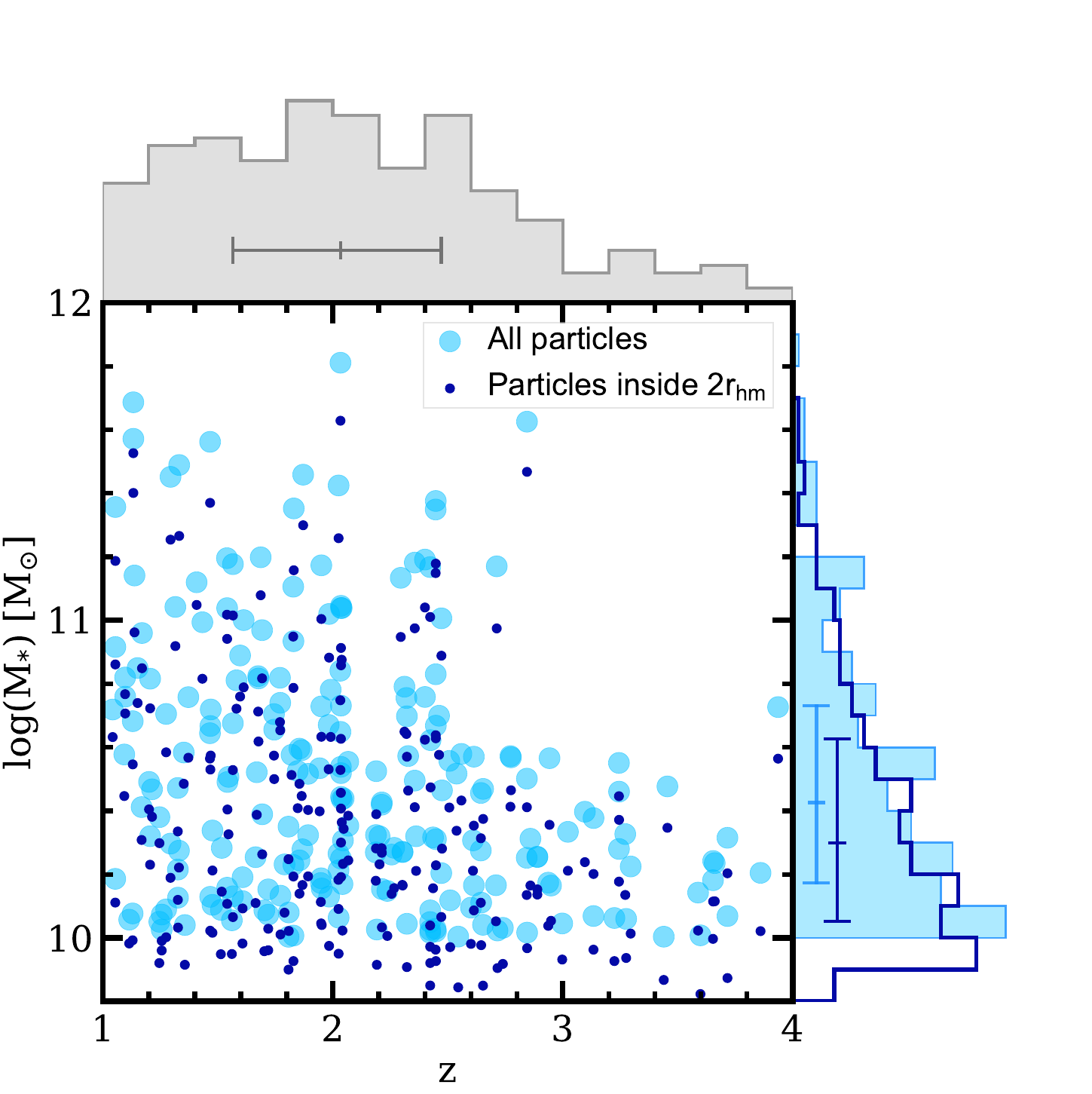}
\caption{Stellar mass vs$.$ redshift plot for galaxies in our sample, with values extracted from the Illustris database. We show the stellar mass when considering all the particles in the galaxy (total stellar masses) as light blue circles and the stellar mass obtained when only particles inside twice the stellar half-mass radius as small dark blue dots. Histograms of stellar masses and redshifts are shown at the top and on the right, with the median and quartiles marked. \label{fig:logM_vs_z}}
\end{figure}

Fig.~\ref{fig:logM_vs_z} shows the stellar mass vs$.$ redshift plot for our galaxies. Two different values of stellar masses are shown for each galaxy: considering all particles in the galaxy or only particles inside twice the stellar half-mass radius. The median values and quartiles for these mass measurements, in log(M$_{\ast}$/M$_{\odot}$), are $10.43_{10.17}^{10.73}$ and $10.30_{10.05}^{10.63}$, respectively, and $2.03_{1.57}^{2.47}$ for the redshifts. 

\subsection{Ground-truth Physical Properties of each Galaxy}\label{subsec:sfh_ground}

We use the Illustris database to extract the information for all the simulated particles belonging to a galaxy. In particular, we extract total stellar masses and SFRs of the whole galaxy. These total stellar masses and SFRs values are the ones shown in Fig.~\ref{fig:fig_sample}.

We build ground-truth galaxy SFHs from the information of the individual particles that belong to each galaxy. The SFH can be computed by first loading from the database the stellar particles in the galaxy and then making a histogram of their formation ages in lookback time. We assume time bins for the formation ages of 25~Myr up to a formation age of 1~Gyr, and 250~Myr afterwards. Subsequently, the stellar mass formed per time interval is calculated by summing the masses of the stellar particles formed in each time bin. We will also use in the following sections an additional SFR estimation for each galaxy obtained from the SFH and calculated by loading the gas particles in the galaxy for the snapshot corresponding to the observed redshift and adding together their instantaneous SFR values. Since these ground-truth SFHs will be compared with those derived from the SED-fitting in Section~\ref{sec:evaluation}, in order to make the comparison more fair, we only consider stellar and gas particles which are closer to the galaxy center than the radius of the photometric aperture. Still, our SFHs will be naturally affected by the limitation inherent to detection and photometry procedures.

\newpage
\section{ESTIMATION OF THE SFH FROM 2D SED FITTING} \label{sec:SPSmodel}

Our aim is to study the earliest formation phases of nearby massive galaxies by analyzing in detail the formation history and location of the star formation in massive galaxy progenitors at high redshift. For that purpose, our approach consists in using broad-band data covering the (observed-frame) optical-to-mid-infrared spectral range provided by HST and JWST. In this paper, we assess the robustness of the results derived from this type of realistic but naturally simplified analysis using simulated imaging data from Illustris (see  Section~\ref{sec:data}). As described in the previous section, we build SEDs for each source for both the integrated emission as well as in a 2D grid. In this section, we describe the derivation of integrated SFHs for each galaxy based on SPS modeling for these SEDs. We then analyze how successfully we can recover the ground-truth SFHs when only broad-band HST and JWST photometric data is used.

\subsection{Stellar Populations Synthesis Modeling} \label{subsec:subSPSmodel}

The integrated and grid SEDs are compared with stellar populations models of Starburst99 (\citealt{1999ApJS..123....3L}), assuming a \cite{2001MNRAS.322..231K} IMF and a \citet{2000ApJ...533..682C} attenuation law. To perform these fits, we use the \textit{synthesizer} code described in \citet{2003MNRAS.338..508P,2008ApJ...675..234P}. This code combines the emission of both stars and gas, compares the models with the observed data, and returns the model that best fits the data by performing a $\chi^{2}$ minimization. 

We assume a double-burst delayed exponential SFH, with each burst described by $SFR(t) \propto t \cdot e^{-t/\tau}$ for $t>0$ up to $t=t_\mathrm{burst}$, being $\tau$ the star formation timescale and $t_{burst}$ the age of the burst. The age of the burst must be understood as the time passed between the age of the Universe when the galaxy started to form stars in that burst and the time corresponding to the observed redshift. The SFH form was chosen after some testing which indicated that 2 bursts following a delayed-exponential SFH instead of other simpler parametrizations (e.g., one single exponential burst) as a more adequate parametrization to successfully reproduce the ground-truth SFHs given by Illustris. Several tests were performed assuming only one time-delayed exponential with different values for $\tau$ and the minimum age of the population, but all of them resulted in the 2D-SED fits selecting too young (recent) ages for the models which were unable to recover the ground-truth galaxy SFH. We will discuss this further in Section~\ref{sec:evaluation}.

The free parameters in our SED fits are the age ($t$), star-formation timescale ($\tau$), extinction ($A_V$), and metallicity ($Z$) for the two stellar populations, in addition to the burst strength ($b$). This burst strength describes the fraction of the total stellar mass that has been created by the most recent burst. We consider an old burst as the one occurring first in the galaxy formation history and a young burst as the one occurring closer to the age of the Universe corresponding to the observed redshift. The stellar mass (M$_{\ast}$) is derived from the SED fits by normalizing the best-fitting model (which provides mass-to-light ratios at all wavelengths) to the observed photometry.

In order to improve the resemblance of the estimated SFHs to the ground-truth SFHs built from simulated particles in the simulation, we found that the allowed age ranges for the old and young stellar populations must depend on the galaxy redshift. In other words, the age frontier between the two stellar populations is an important  parameter to set {\it a priori} based on the galaxy redshift. We thus tested the dependence of our results on this age separation limit (age$_{\text{lim}}$ in Table~\ref{tab:SEDparameters}), considering values as a function of the age of the Universe for a given redshift. The most accurate results in our analysis (considering the mass-fraction formation ages discussed in Section~\ref{sec:evaluation}) are obtained when we impose an age limit between the 2 populations equal to 40\% of the age of the Universe for galaxy redshifts at $1 < z < 2$  and to 50\% of the age of the Universe for $2 < z < 4$ (i.e., 40\% of the age of the Universe for the young population and 60\% for the old population when $1 < z < 2$, and 50\% of the age of the Universe for both populations when $2 < z < 4$). 

Similarly, we tested different values for the star formation timescales of both bursts and found the best results are obtained when setting $200 < \tau_{\text{old}} < 1000$\,Myr, and $\tau_{\text{young}} =$\,10~Myr. Table~\ref{tab:SEDparameters} shows the ranges within which each free parameter is allowed to vary in the SED-fittings of this work. 

It is impossible to present the results of the full set of tests performed to explore the parameter space here, so instead we focus on those that yield the best match and will be the most useful for upcoming JWST observations.

The metallicity is allowed to adopt three different values: Z/Z$_{\odot}$ = 0.2, 0.4, and 1. These values are expected for our sample according to the mass-metallicity relationship at the considered redshifts, at which this relationship shows lower metallicities than locally for a given mass (\citealt{2006ApJ...644..813E}; \citealt{2008A&A...488..463M}; \citealt{{2009MNRAS.398.1915M}}; \citealt{2011ApJ...730..137Z}). Regarding the dust attenuation, we limit the attenuation range to values within $0<A_{\text{V,old}}<1$~mag for the old population, and within $0<A_{\text{V,young}}<2$ for the young population.

\begin{deluxetable}{cccc}[t]
\tablenum{3}
\tabletypesize{\footnotesize}
\tablecaption{Free parameters and their ranges in the SED fitting for a double delayed-exponential SFH \label{tab:SEDparameters}}
\tablehead{
\colhead{Parameters}  & \colhead{Values/Range} & \colhead{\phantom{ss}Units\phantom{ss}}& \colhead{Step} \\[-0.55cm]
}
\startdata
$t_{\text{young}}$  &  \phantom{suss.}0.1 - age$_{\text{lim}}$\tablenotemark{\scriptsize a} & Gyr &  discrete\tablenotemark{\scriptsize b}\\
$\tau_{\text{young}}$ &  10  & Myr &  --\\
$A_{\text{V,young}}$  & 0 - 2  & mag &  0.1 mag\\
\hline
$t_{\text{old}}$ &  \phantom{ssssi}age$_{\text{lim}}$ - age$_{\text{Univ,z}}$\tablenotemark{\scriptsize c} & Gyr &  discrete\tablenotemark{\scriptsize b}\\
$\tau_{\text{old}}$  &  \phantom{i.}200 - 1000 & Myr &  0.1 dex\\
$A_{\text{V,old}}$ &  0 - 1  & mag &  0.1 mag\\
\hline
$Z_{\text{young}}$ \& $Z_{\text{old}}$ &  [0.2, 0.4, 1] \phantom{.} & Z/Z$_{\odot}$ &  discrete\\
$b$ &  0.01 - 1 \phantom{s..} & - &  0.01 \\
\enddata 
\vspace{0.2cm}\scriptsize{\textbf{Notes:}}
\vspace{-0.1cm}\tablenotetext{ a}{\footnotesize The age separation limit between the young and old populations, is measured (backwards) from the redshift of observation of the galaxy. This value depends on the redshift of the galaxy: it is set to 40\% of the age of the Universe for \mbox{1 $<$ z $<$ 2} and  to  50\%  of the age of the Universe for \mbox{2 $<$ z $<$ 4}.}
\vspace{-0.2cm}\tablenotetext{ b}{\footnotesize We use all the discrete values for the ages given by the SB99 models within the allowed range.} 
\vspace{-0.2cm}\tablenotetext{ c}{\footnotesize The maximum allowed value for the age of the old population is the age of the Universe at the galaxy redshift.} 
\vspace{-0.8cm}
\end{deluxetable}

\begin{figure}[t]
\begin{subfigure}
\centering
\includegraphics[scale=0.37, trim=2 61 2 0,clip]{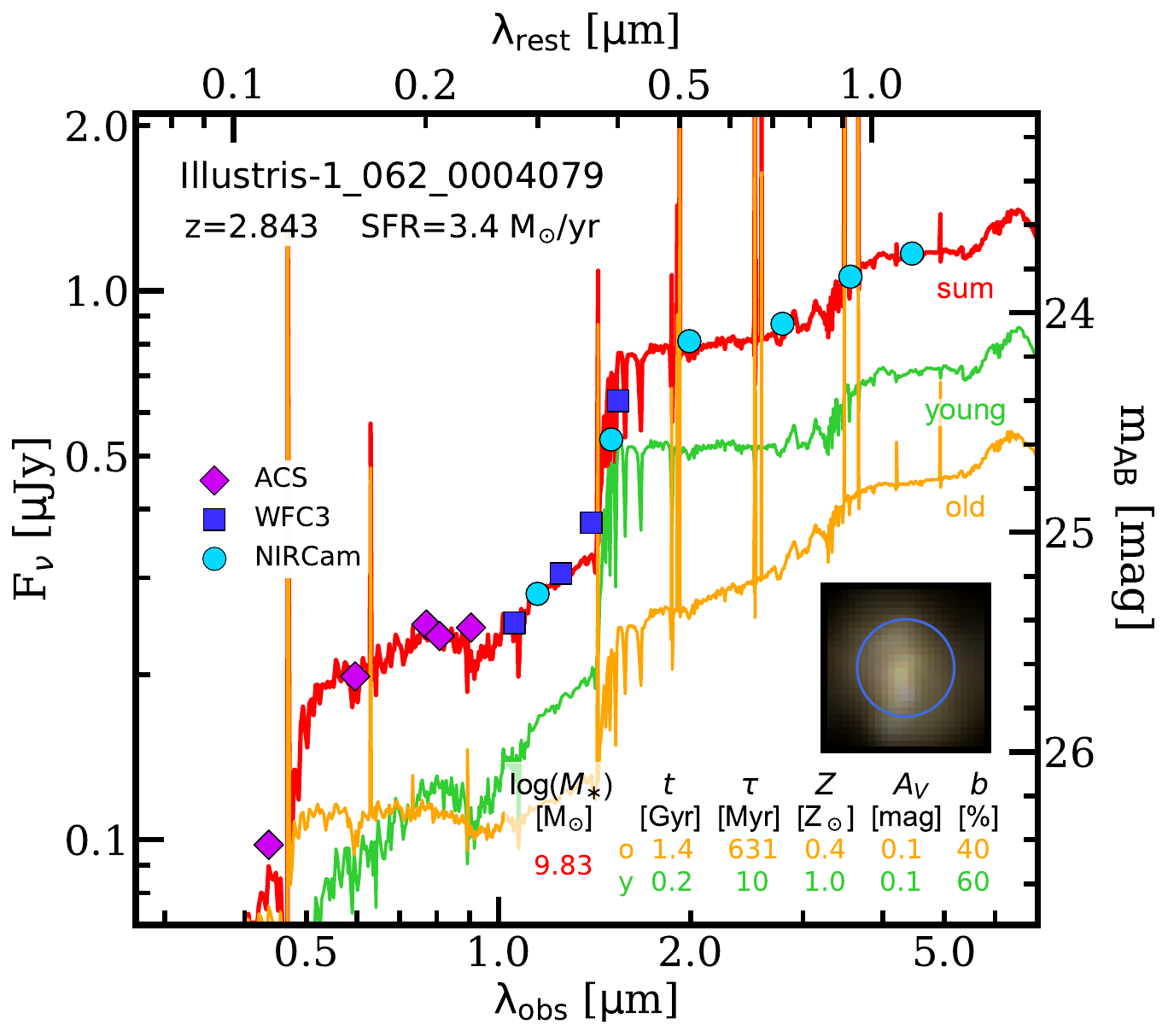}
\end{subfigure}
\begin{subfigure}
\centering
\includegraphics[scale=0.3695, trim=-1.5 0 1.6 14,clip]{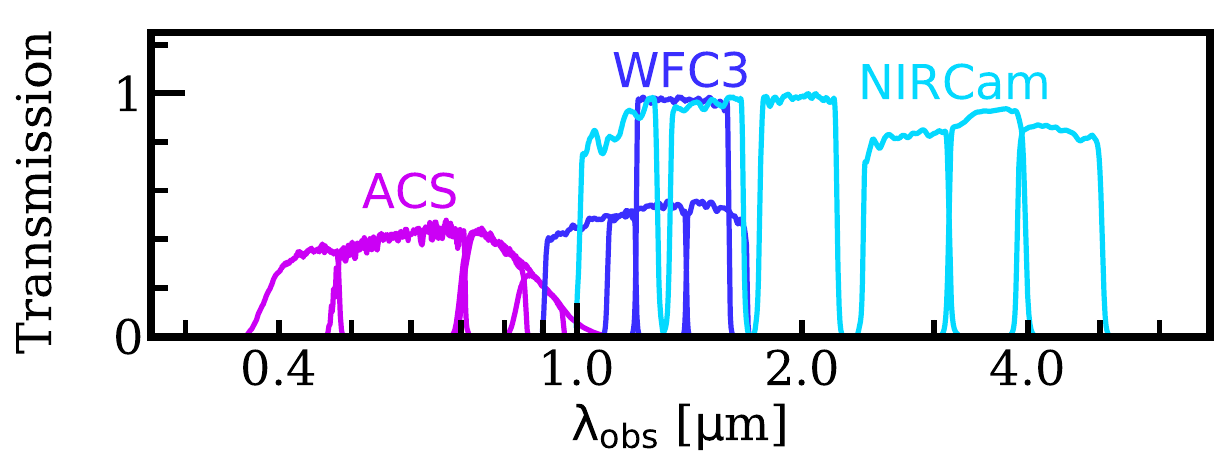}
\end{subfigure}
\caption{Example of one integrated SED fit for one of our galaxies: Illustris-1\_062\_0004079 at $z = 2.843$. This galaxy code stands for galaxy id 4079 and snapshot 62 ($z=2.73$) in the Illustris-1 simulation. The best fit is shown as a red line. We also show the models corresponding to the old (green) and young (orange) stellar populations. The transmission curves of the filters have been included at the bottom of the figure. The best-fit stellar parameters for the old and young populations are also given. An RGB image (size $1.5\arcsec \times 1.5\arcsec$) with the integrated aperture is shown as an inset.  
\label{fig:SED_ex}}
\end{figure}

Fig.~\ref{fig:SED_ex} shows an example of an integrated SED fit and the values of the model that best fits our data. For this particular galaxy, both the old and young stellar populations contribute at the 40\%-60\% level throughout the full spectral range. This analysis is performed for the 221 galaxies: both for their integrated measurements and their 2D emission (given by the cells in the grid).

To estimate uncertainties in the derived stellar population properties, Monte Carlo (MC) simulations are performed in \textit{synthesizer} by allowing the photometric data to normally vary within their photometric uncertainty (without correlation), and then repeating the fit again for 300 resampled SEDs \citep[more details in, e.g.,][]{2016MNRAS.457.3743D}. This results in 300 sets of solutions for each SED which also provide us with information about the typical degeneracies present in these kinds of studies (e.g., age-metallicity or $\tau$-age degeneracy).

\begin{figure}[t]
\centering
\includegraphics[width=0.95\columnwidth, trim=0 3 0 -14.5,clip]{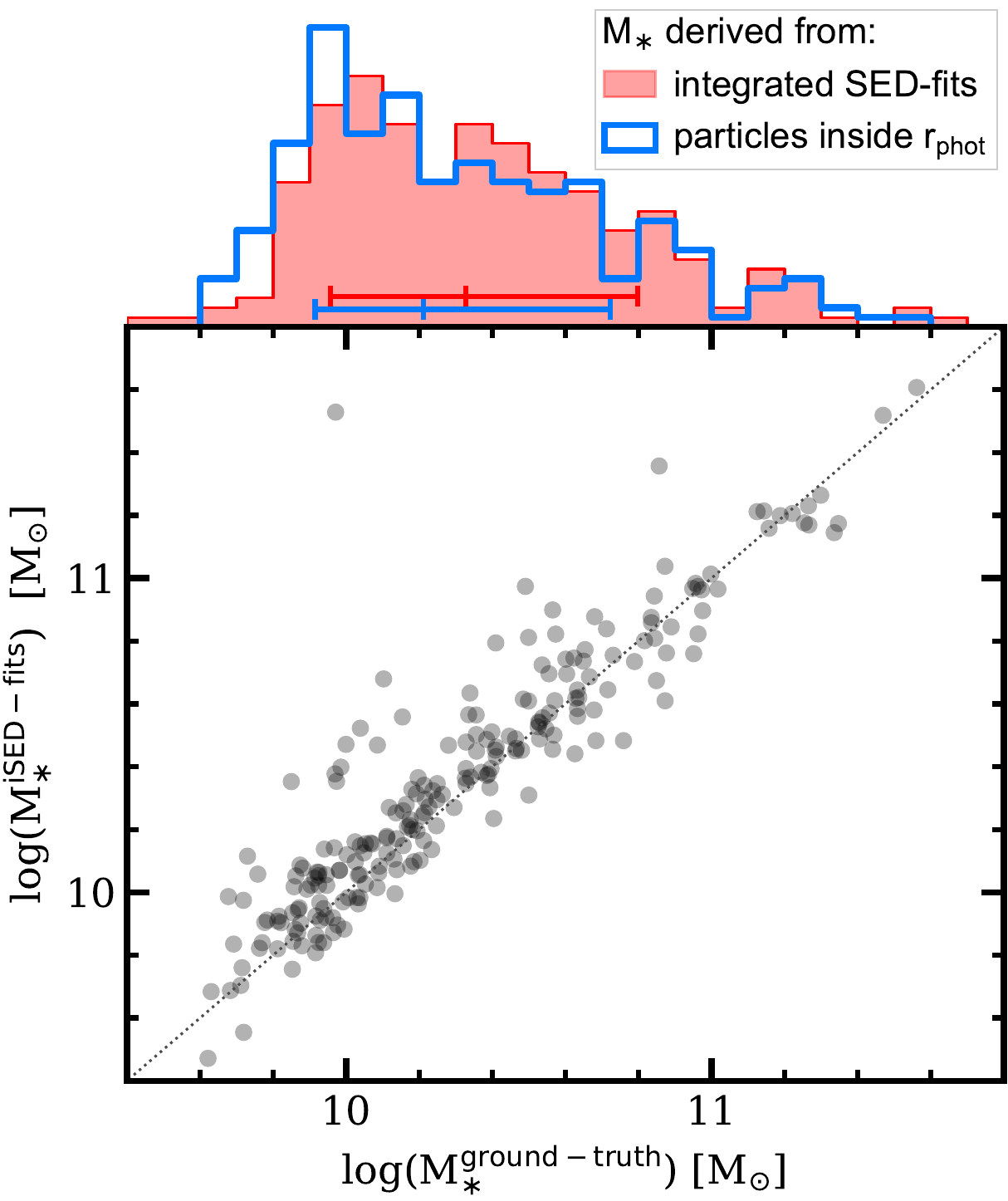}
\caption{Stellar masses for the $1 < z < 4$ sample derived from the integrated SED-fits vs$.$ their ground-truth values calculated from the simulated stellar particles belonging to each galaxy (inside a sphere with the same radius as that of the integrated photometric aperture). The one-to-one relation is shown with the dotted black line. The histograms for both distributions are shown in different colors, with the median and the 68\% intervals as horizontal segments.
\label{fig:iM_hist}}
\end{figure}

Fig.~\ref{fig:iM_hist} shows the stellar masses for our $1<z<4$ sample derived from the integrated SED-fits vs$.$ their ground-truth values and how the one-to-one relation is recovered, with a median offset of 0.04 dex and a scatter of 0.2 dex. Each of the SPS-derived galaxy masses has been calculated as the median stellar mass provided by the integrated SED fits. The ground-truth masses have been calculated from the database by considering only the simulated particles of each galaxy closer to the galaxy center than the radius of the integrated photometric aperture. In both cases, stellar masses correspond to the current mass of stars at the redshift of observation (i.e., calculated after taking into account the time-dependent mass-loss and fraction of remnants as a function of time). The histograms for both distributions are also shown. We notice our SPS-derived stellar masses slightly overestimate the ground-truth masses, although the differences are small: \mbox{log(M$_{\ast}$/M$_{\odot}) = 10.33_{\phantom{1}9.96}^{10.80}$} for the integrated SED-fits vs$.$ $10.21^{10.72}_{\phantom{1}9.92}$ for the stellar particles (median and 68\% confidence interval).

Several effects could be responsible for the dispersion observed around the one-to-one relation between our integrated vs$.$ ground-truth masses. One of these effects is due to the difference between the volume of the sphere used to calculate the ground-truth masses and that of the cylinder along the line of sight which would enclose all the stars contributing to photometry. This makes that, depending on the shape of the galaxy, when measuring photometry we may include the light from stellar particles in the line of sight located at large galactocentric distances (not included in spherical aperture used to calculate the ground-truth galaxy mass). In fact, this is what happens for the galaxy with an integrated mass that differs in $\sim$1.5$\,$dex in Fig.~\ref{fig:iM_hist}, which is being outshined by an ultra-massive galaxy of $\log($M$_{\ast}^{\text{ground-truth}}$/M$_{\odot}) = 11.47$ located at only 0.1$\,\arcsec$ from its center. Another effect which can cause dispersion in the recovered integrated mass is the presence of nearby neighbors and their identification in the segmentation map. Even though the light of these neighbors inside the photometric aperture is usually (but not perfectly) blocked by the segmentation map, so is the light of the considered galaxy inside this masked region. However, for the calculation of the ground-truth galaxy mass, we do take into account these stellar particles (from the considered galaxy) located in this region of the sphere.

\subsection{Estimating SFHs from HST+JWST Photometry}\label{subsec:phot_sfhs}

\begin{figure*}[t]
\centering
\includegraphics[width=0.885\textwidth, trim=2 52 4 9,clip]{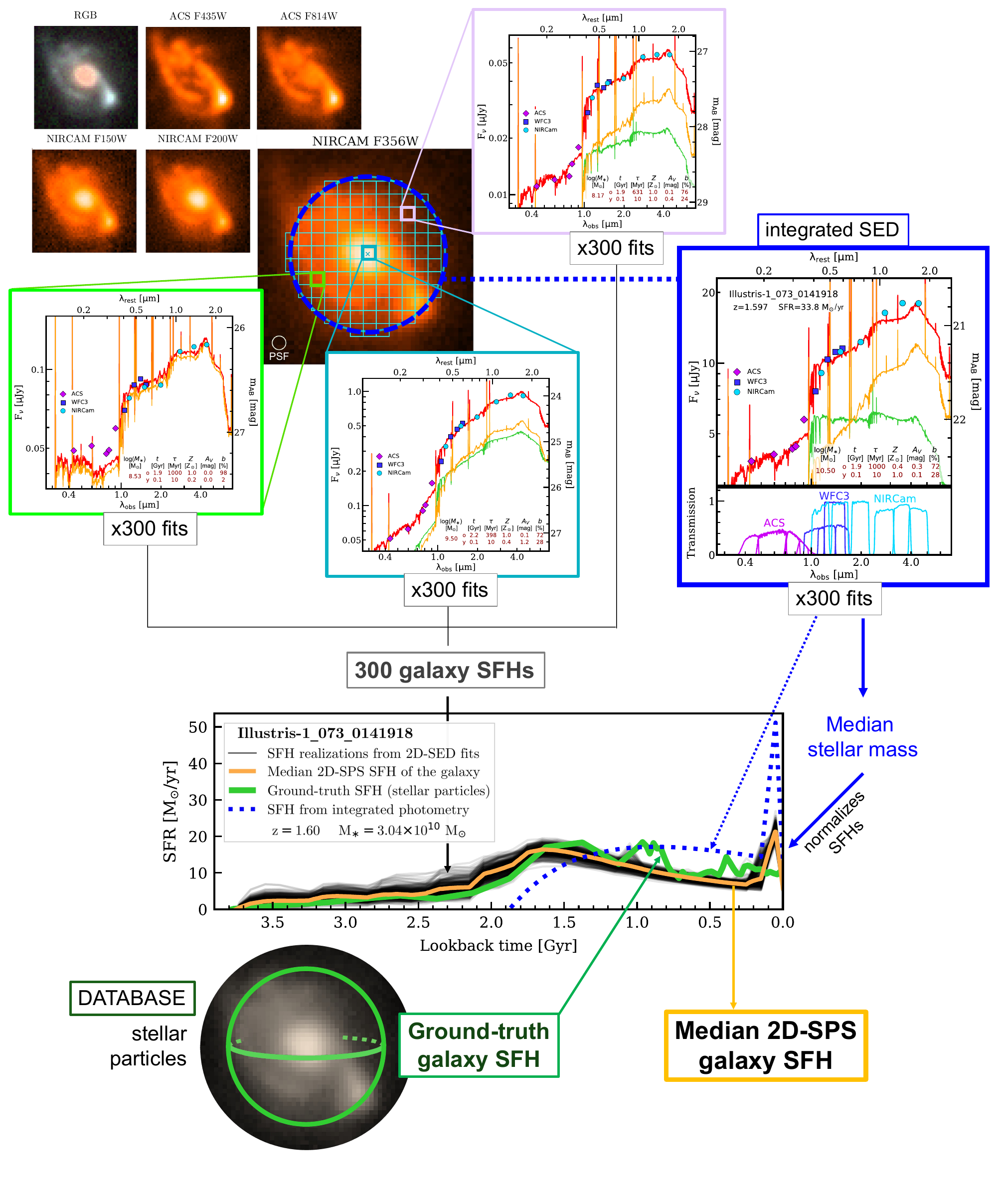}
\caption{Schematic diagram of the methodology followed to obtain the median SFH for a galaxy from its 2D-SPS analysis. First, we show an RGB image of the galaxy and some of the HST+JWST broad-band images (in Table~\ref{tab:filters}) processed to imitate CANDELS and CEERS observations for HST and JWST filters, respectively. We measure the 2D photometry inside a grid (in cyan) with cell size equal to the spatial resolution element, in addition to the integrated photometry (blue aperture). As an example, we show the SEDs measured for three regions of the grid: the center of the galaxy (teal cell), an arm region (light green), and a diffuse emission zone (pink). To estimate the uncertainties and degeneracies in the derived stellar populations, each SED is fitted 300 times by performing MC simulations. We add the individual SFHs inferred from all the grid regions to obtain the SFH for the whole galaxy. The black thin SFHs in the bottom subfigure show the SFHs for the whole galaxy created by accounting for the uncertainties in the SFHs of each grid region, where the median galaxy SFH from this 2D-SPS method is shown in yellow. The blue dotted SFH is the galaxy SFH inferred from the integrated photometry. The ground-truth SFH of the galaxy, given by the stellar particles belonging to the galaxy, is the green solid SFH.\label{fig:method}}
\end{figure*}

Integrated SFHs are built for each galaxy from the 2D SED-fits, and compared with the ground-truth SFHs extracted from the Illustris database (see Section~\ref{subsec:sfh_ground}). We build 300 integrated SFHs for each galaxy using the 2D stellar population fits, which includes a MC method to estimate uncertainties and degeneracies. We first create one SFH for each resolution element in the galaxy grid. Subsequently, one global SFH is calculated by adding together all these SFHs in the grid. This procedure is then repeated 300 times, one for each of the MC particles presented in the previous section. By combining and adding the 300 SED-fits for each cell in the grid, we obtain 300 independent estimations of the integrated SFH for a given galaxy. 

The resulting SFHs are smoothed using a 100 Myr square kernel. Then, we normalize these 300 galaxy SFHs with the median stellar mass of the galaxy derived from the 300 integrated SED-fits. Unless otherwise stated, these galaxy stellar masses refer to the galaxy masses at the time of observation without including remnants or yields. Finally, we calculate the median SFH of the galaxy from these 300 normalized SFHs. 

In Fig.~\ref{fig:method}, we summarize the methodology followed for a galaxy to obtain this median 2D-SPS galaxy SFH from its 2D photometry measurements. We start from the HST+JWST data (RGB image and some individual bands shown on the figure), in our example, a galaxy presenting a red center which resembles a protobulge and what seems like a blue disk with some spiral structure. We measure integrated and 2D photometry in a grid. We show SEDs and their stellar population modeling results for some representative grid regions in the center, blue arm on the bottom left and a diffuse emission zone on the top right. We note the difference in color shown in the SEDs between the center and the disk, and how the integrated SED resembles more that of the galaxy center rather than the disk regions. In this regard, considering 2D photometry in our analysis facilitates the estimation of SFHs since small parts of a galaxy are expected to have more simple forms than the entire galaxy, whose integrated photometry is dominated by the regions with highest intensities, i.e., they overshine fainter regions located in the outskirts for this example galaxy. The individual SFHs for each grid region are added to obtain an integrated SFH. In this example, we demonstrate how the derived SFHs with our method nicely reproduce the ground-truth SFH, significantly improving what can be obtained by analyzing the integrated SED. In the following section, we compare our derived SFHs from our 2D analysis of broad-band imaging data with the reference SFHs built using the Illustris-1 database in a statistical way.

\newpage
\section{VALIDATION OF THE METHOD} \label{sec:evaluation}

In this section, we evaluate the levels of success of our method in recovering the SFHs of massive galaxies at high redshift. In particular, in this paper we are mainly interested in the earliest phases of massive galaxy formation. Therefore, we compare our results based on the analysis of broad-band HST and JWST images with the ground-truth provided by the Illustris database for full galaxies. We will analyze spatially resolved stellar population properties in a future paper, here we concentrate on the results about the SFHs.

\subsection{Characterization of the Earliest Phases in the Formation of Massive Galaxies}

In order to characterize when massive galaxies start their formation, we calculate the formation times and redshifts when galaxies have formed a given (small) fraction of their total stellar mass. In particular, we discuss here the formation times when the first 5\%, 10\% and 25\% of the stars in each galaxy and the entire sample were formed. We define $t_{\text{k}}$ as the formation time (measured from the galaxy redshift) at which a galaxy formed the k$\%$ of its total stellar mass present at the galaxy redshift.  These mass-fraction formation times, $t_{\text{k}}$, can be directly computed by integrating the 2D-SED-derived galaxy SFHs over cosmic time and are similar to other parameters used in recent literature \citep[see, e.g.,][]{2022arXiv220402414J}. We remark that $t_{\text{k}}$ depend on both $t_{\text{0}}$ and $\tau$, so they might somehow alleviate possible degeneracies between ages and timescales. We also calculate mass-weighted ages, $t_{\text{mw}}$, from the SFHs. To test their robustness, these $t_{\text{5}}$, $t_{\text{10}}$, $t_{\text{25}}$, and $t_{\text{mw}}$ extracted from the SFHs derived with our 2D-SPS method are compared with their ground-truth values, which are obtained from the SFHs of the simulation stellar particles following the same procedure and enclosed by the photometric aperture. 

\begin{figure*}[ht!]
\centering
\includegraphics[width=0.9595\textwidth, trim=0 6 0 6, clip]{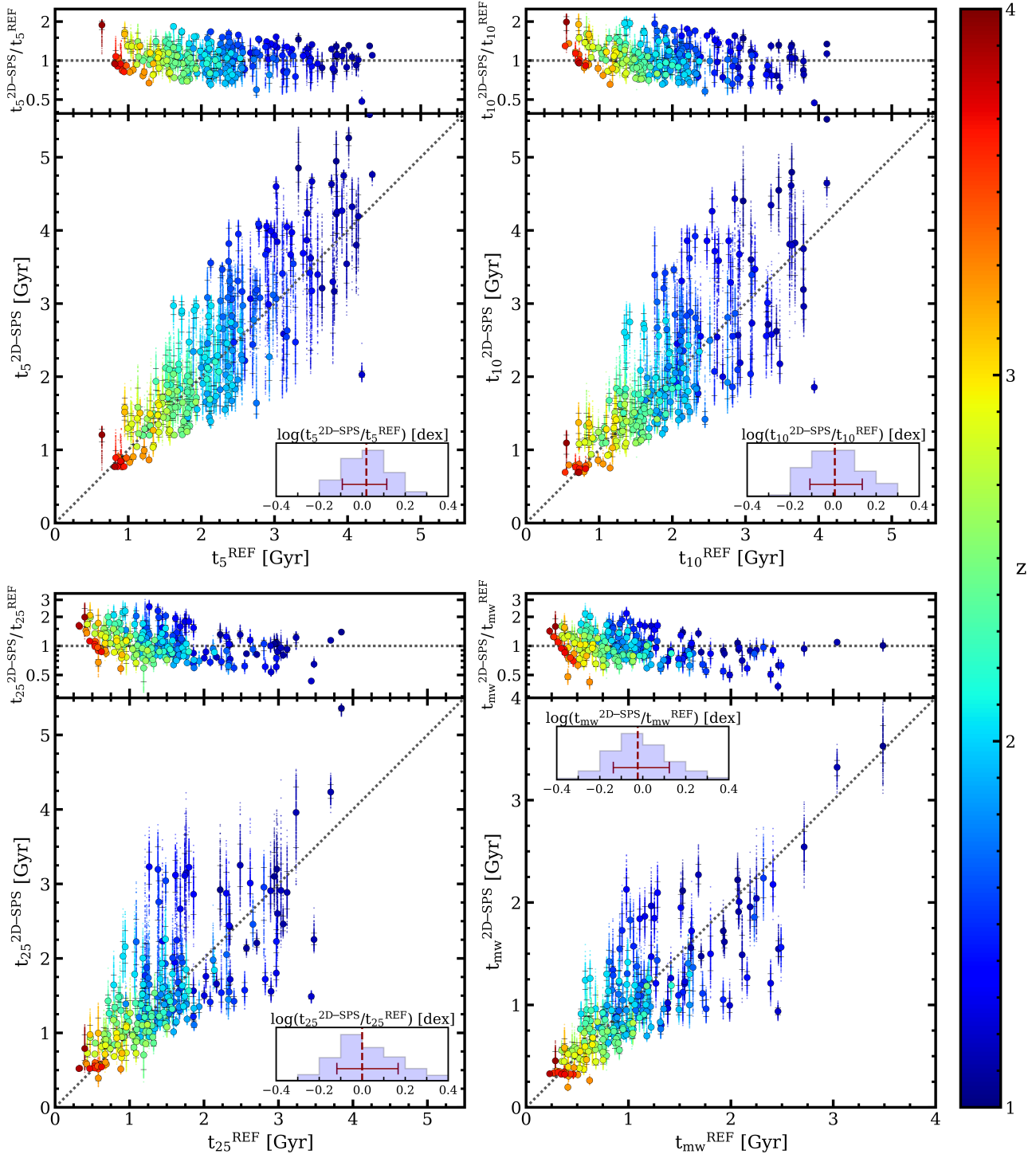}
\caption{Comparison between the mass-fraction formation times (measured backwards from the redshift of observation of the galaxy) of our 2D-SPS method and their ground-truth values: $t_{\text{5}}$ (upper-left), $t_{\text{10}}$ (upper-right), $t_{\text{25}}$ (lower-left) and $t_{\text{mw}}$ (lower-right) calculated from our 2D-SPS-derived SFHs versus their reference values calculated from the SFHs built from the simulated stellar particles in galaxies. Each galaxy is represented by 300 vertically spread points, which correspond to the mass-fraction formation times of the 300 2D-SPS SFHs, built from the 300 MC particles in each resolution element in the grid. The median of these values for each galaxy is shown as bigger circles and error bars represent the standard deviation of these values (68\%  interval). All points are color-coded by galaxy redshift. On the top of each panel, we show the ratio between the $t_{\text{k}}$ of our 2D-SPS method and their ground-truth values as a function of the latter. As an inset, we include the offsets of the galaxies in the sample. \label{fig:final_test_4050_taufixed10_sb99_3x3}}
\end{figure*}

Fig.~\ref{fig:final_test_4050_taufixed10_sb99_3x3} shows $t_{\text{5}}$, $t_{\text{10}}$, $t_{\text{25}}$, and $t_{\text{mw}}$  calculated from the SFHs derived with 2D stellar population modeling of broad-band data versus their reference values. For each galaxy, we show 300 values of these quantities (vertically spread out), which correspond to the 300 MC particles or 2D-SPS galaxy SFH as described in Section~\ref{subsec:phot_sfhs}. The black dotted line shows the one-to-one relation between values derived from our 2D-SPS method (output) and those from the simulated particles (ground-truth). As explained in Section~\ref{subsec:subSPSmodel}, some {\it a priori} parameters of our method (e.g$.$ Table~\ref{tab:SEDparameters}) have been optimized to reproduce this one-to-one relation for these four mass-fraction formation times. In general, we find our ages are consistent with this relation at all redshifts. As an inset in each panel, we show the histogram of relative differences between median and ground-truth values for all galaxies. We find our $t_{\text{5}}$, $t_{\text{10}}$, $t_{\text{25}}$, and $t_{\text{mw}}$  values are consistent with the ground-truth values with a median (relative) offset of $+71$ ($+4.4$\%), $+$16 ($+1.8$\%), -2 (-0.1\%), and -50 Myr ($+5.2$\%), and a scatter (68\% interval) of 0.33, 0.31, 0.27, and 0.17 Gyr, respectively. As commented in Section~\ref{subsec:subSPSmodel}, when we consider only a single-population SFH roughly spanning the whole range of parameters ($t$, $\tau$, $A_{V}$, $Z$) of the two-population fitting, too young ages are assigned in the SED-fits, which leads to an underestimation of all the ground-truth formation times. Thus, when we assume a SFH given by only one population instead of two, these median offsets are never better than -0.9, -0.8, -0.5, and -0.4$\,$Gyr for $t_{\text{5}}$, $t_{\text{10}}$, $t_{\text{25}}$, and $t_{\text{mw}}$, respectively.

We have included on the top of each panel the ratio between the $t_{\text{k}}$ of our 2D-SPS method and their ground-truth values. In general, we do not see any systematic effect as a function of redshift, but we do find our $t_{\text{25}}^{\text{2D-SPS}}$ and $t_{\text{mw}}^{\text{2D-SPS}}$ tend to underestimate ground-truth in some galaxies as the ground-truth $t_{\text{k}}$ increases: $t_{\text{25}}$ ($t_{\text{mw}}$) presents a median systematic offset of only $\sim 5$\% ($-2$\%) with respect to ground-truth for lookback times younger than 2 (1.5)$\,$Gyr, but $\sim -13$\% ($-16$\%) for older lookback times. We do not see this behaviour in the case of $t_{\text{5}}$ and $t_{\text{10}}$, which present a median systematic offset of less than $\sim6$\% for ages younger and older than 2$\,$Gyr. A possible interpretation for this bias in $t_{\text{25}}$ and $t_{\text{mw}}$ observed for some low-redshift galaxies could have its origin in the double-burst SFH model assumed for the 2D-SPS analysis. This model causes the galaxy SFH to usually rise quickly in later epochs due to the presence of the young population burst (see the median 2D-SPS galaxy SFH in Fig.~\ref{fig:method}). As a consequence of this late and rapid increase in the SFR, the middle of the SFH necessarily presents lower SFR in order to reproduce the given final light or mass, which would produce a bias when deriving those formation times of the galaxy that are more sensitive to this middle part of the SFH, i.e., $t_{\text{25}}$ and $t_{\text{mw}}$. Since low-redshift galaxies have a more extended and possibly more complex SFH than galaxies at high-redshift, this effect would be more noticeable in them.

Regarding the precision of our method, we find the scatter values are relatively small and similar for all $t_{\text{k}}$, but we observe the scatter tends to increase from $t_{\text{5}}$ to $t_{\text{25}}$. Taking all this into account, we conclude that our 2D-SPS method successfully recovers $t_{\text{5}}$, $t_{\text{10}}$, $t_{\text{25}}$ and $t_{\text{mw}}$ for $1<z<4$ massive galaxies with a $\sim$25\% uncertainty and small $<$3\% systematic effects.

\section{EXPECTATIONS FOR THE DERIVATION OF THE SFH OF $z>1$ MASSIVE GALAXIES WITH HST+JWST DATA}\label{sec:results}

The aim of this section is to discuss when the early stages of stellar mass assembly took place in very massive galaxies. To do this, we apply stellar population synthesis in 2D on our galaxy sample at $1<z<4$ and we compare the statistical results on the SFHs for the sample with the ground-truth values inferred from the simulation stellar particles. This comparison gives us information about the limitations and observational biases we will encounter when using this method on galaxy samples constructed with real data.

\subsection{When Did Massive Galaxies Begin to Form?}

In Section~\ref{sec:evaluation}, we showed that our 2D-SPS method successfully recovers $t_{\text{5}}$, $t_{\text{10}}$, and $t_{\text{25}}$ for our galaxy sample. These quantities represent the lookback times at which galaxies reach 5\%, 10\%, 25\% of the stellar mass formed, respectively, and can be used to estimate the beginning of star formation in galaxies. To address the question of when the population of very massive galaxies began to form, our approach consists in calculating $t_{\text{5}}$, $t_{\text{10}}$, and $t_{\text{25}}$ from a median SFH of the sample built from the 2D-SED fits. At this point, we should remind the reader that our sample consists of galaxies at $1 < z < 4$ with M$_{\ast} > 10^{10}$ M$_{\odot}$, all having very massive descendants (M$_{\ast} > 10^{11}$ M$_{\odot}$) at $z=0$. Hence, we expect the median SFH from our 2D-SPS modeling to resemble the median SFH of very massive galaxies at $z=0$ over the redshift interval of our sample (where both SFHs overlap). We say \textit{resemble} because our sample limitations might have implications on the derived SFHs, i.e., we have to consider the systematic effects introduced by not taking into account lower mass galaxies and by the relative number of galaxies selected at different redshifts in our full interval.

Since the final goal of our 2D-SPS method is to be applied on real galaxy samples at high redshift as soon as JWST data are available, we need to check first whether our results from the 2D-SED fits regarding the first episodes of stellar assembly in very massive galaxies are compatible with the ground-truth of the simulation. To address this issue, we build two additional samples of galaxies at $z=0$ in the simulation to be compared with our main $1<z<4$ galaxy sample. For these initial three samples of galaxies, we build the typical (median) ground-truth SFH from the simulated particles in their galaxies, which will be compared with the typical SFH from the 2D-SPS analysis on the $1<z<4$ studied sample of massive progenitors. The  samples of galaxies in the Illustris-1 simulation considered in this section are:

\begin{enumerate}[(1)]
    \item Our main galaxy sample at $1 < z < 4$ used for the 2D-SPS analysis. It consists of the 221 (out of the 248) M$_{\ast} > 10^{10}$ M$_{\odot}$ progenitors at $1 < z < 4$  in the \citetalias{2017MNRAS.468..207S} images with a very massive descendant at $z=0$, and which have not been discarded during the analysis procedure (see Section~\ref{subsec:sample}). This is the ``high-redshift sample of (massive) progenitors" or ``main sample", hereafter.
    \item The descendants at \mbox{$z=0$} of our main high-redshift galaxy sample, all of them with M$_{\ast} > 10^{11}$ M$_{\odot}$ (132 descendants). This sample is a subset of the whole population of very massive galaxies at $z=0$. 
    \item The whole population of very massive galaxies (M$_{\ast} > 10^{11}$ M$_{\odot}$) at $z=0$. This sample is composed by 856 galaxies. 
\end{enumerate}

Although the main sample of this study is composed of massive progenitors of local M$_{\ast} > 10^{11}\,$M$_{\odot}$ galaxies, our selection criteria could not be applied as such to real observations where the estimation of the final mass at $z = 0$ of observed high-redshift galaxies is not easy without additional assumptions. To study the impact of this, we apply the same 2D-SPS method described for our main sample of $1<z<4$ progenitors to the following fourth, additional sample:
\begin{enumerate}[(4)]
    \item All M$_{\ast} > 10^{10}\,$M$_{\odot}$ galaxies at $1<z<4$ in the \citetalias{2017MNRAS.468..207S} images, regardless of the stellar mass of their descendants at $z=0$. It consists of the 350 (out of the 388) massive galaxies at $1<z<4$ in the images that are not discarded when the 2D-SPS method is applied to them. This sample is a combination of our main galaxy sample of 221 massive progenitors plus other 129 massive galaxies at $1<z<4$ in the images that do not reach M$_{\ast} > 10^{11}\,$M$_{\odot}$ at $z=0$ for any reason.
\end{enumerate}

The median SFH of this sample of $1<z<4$ massive galaxies built from the 2D-SPS analysis, and also the ground-truth median SFH from the simulated particles in each galaxy, will also be compared to the results of our main sample of $1<z<4$ progenitors.

We remind the reader that the \citetalias{2017MNRAS.468..207S} images only contain a limited number of galaxies from the full Illustris-1 simulation. Thus, the descendants at $z=0$ of the galaxies we have studied through their HST+JWST simulated imaging data are also a limited subset of all the galaxies at $z=0$ in the simulation. Our aim is to see if, by analyzing the SFHs for high-redshift, M$_{\ast} > 10^{10}$\,M$_{\odot}$  progenitor galaxies in the images of $z=0$ very massive galaxies, we can learn when the whole population of very massive galaxies began to form. 

If we consider the masses of all the simulated stellar particles in galaxies, our main $1<z<4$ sample of 221 galaxies accounts for only 28\% of the total stellar mass present in their 132 $z=0$ descendants, of which $1 < z < 2$,  $2 < z < 3$, and $3 < z < 4$ galaxies in the sample would account for 33\%, 23\%, and 8.1\% of the stellar mass in their descendants, respectively. If we also take into account the total stellar masses of less massive progenitors at $1<z<4$ in the images, none of them considered by the mass cutoff, this number raises to 32\% (36\% for galaxies at $1 < z < 2$, 27\% for $2 < z < 3$, and 11\% for $3 < z < 4$). This means that the remaining 68\% of the stellar mass in nearby massive galaxies must be explained by either more recent \textit{in-situ} star formation events or by subsequent mergers at $z<1$. We refer the reader to Section~\ref{subsec:sample} for more information about the main $1<z<4$ galaxy sample and their $z=0$ descendants.

\begin{figure}[t]
\includegraphics[width=\columnwidth, trim=0.7 0 0 -30, clip]{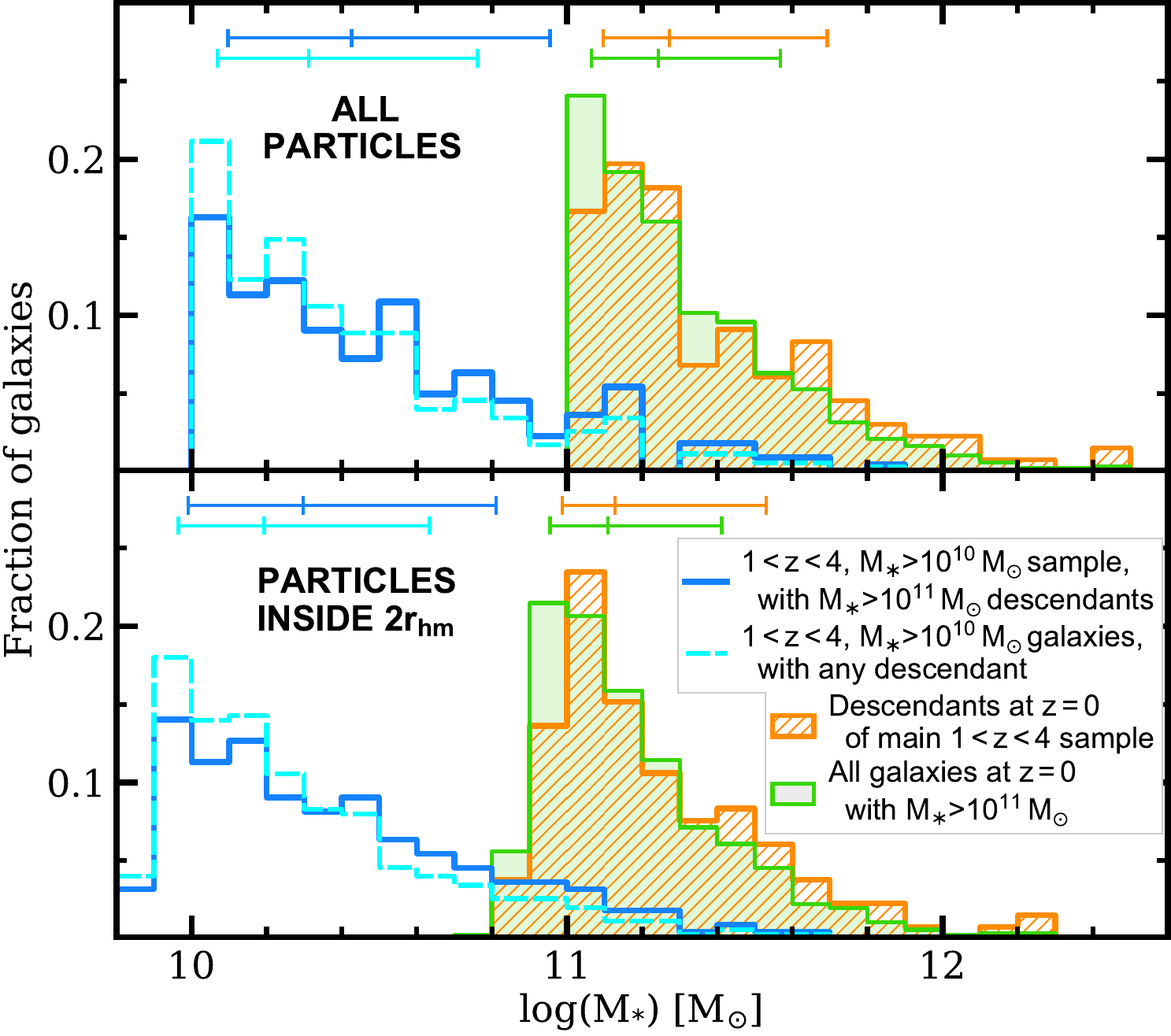}
\caption{Ground-truth stellar mass histograms for the different samples extracted from the Illustris-1 database: when considering all particles in galaxies (top panel) or only particles inside twice the stellar half-mass radius (bottom panel). We show our main 2D-SPS sample of progenitors at $1<z<4$ in red, massive galaxies at $1<z<4$ in cyan, the descendants at $z=0$ of our $1<z<4$ progenitors in orange, and the whole population of very massive galaxies at $z=0$ in dark gray. The median and the 68\% intervals are shown at the top of each panel. 
\label{fig:res_hists}}
\end{figure}

In Fig.~\ref{fig:res_hists} we compare the histograms for the ground-truth stellar masses of galaxies in the four samples, calculated from the simulated particles in the database. On different panels we show the histograms for the stellar masses corresponding to all particles in galaxies and to particles inside 2$\times$r$_{\text{hm}}$. In both cases, the distribution of the stellar masses for our $1<z<4$ sample of massive progenitors resembles that of all massive galaxies at $1<z<4$, although the former is biased towards higher masses: the median and 68\% interval are \mbox{log(M$_{\ast}$/M$_{\odot}$)\,=\,$10.4_{10.1}^{11.0}$} vs$.$ $10.3_{10.1}^{10.8}$ when all particles are considered, and $10.3_{10.0}^{10.8}$ vs$.$ $10.2_{10.0}^{10.6}$ for particles inside 2$\times$r$_{\text{hm}}$, respectively. Regarding the two samples of $z=0$ galaxies, we find that the distribution of masses for the population of very massive galaxies at $z=0$ and for the specific descendants are very similar: log(M$_{\ast}$/M$_{\odot}$)\,=\,$11.3_{11.1}^{11.7}$ vs$.$ $11.2_{11.1}^{11.6}$ (all particles), and $11.1_{11.0}^{11.5}$ vs$.$ $11.1_{11.0}^{11.4}$ (particles inside 2$\times$r$_{\text{hm}}$). As discussed below in this section, this difference in the median masses is a consequence of the mass-cutoff imposed for the selection of the $1<z<4$ sample and will have an impact on the mass-fraction formation times of both samples. In summary, we conclude that our limited sample of progenitors would indeed provide representative results for the full population of massive galaxies at $z=0$. 

In Section~\ref{subsubsec:particles}, we calculate the median ground-truth SFH for these subsets of galaxies using only the simulated particles in them, while in subsection~\ref{subsubsec:SFH_2DSPS} we compare these results with the median SFH of our main high-redshift sample of M$_{\ast}>10^{10}$ M$_{\odot}$ precursors built using our 2D-SPS method. Finally, in subsection ~\ref{subsubsec:varietySFHs} we show the variation of the onset of star formation of massive galaxies.

\subsubsection{Ground-truth SFHs: very Massive $z=0$ Galaxies in Illustris and the $1<z<4$ Sample}\label{subsubsec:particles}

\begin{figure}[t]
\epsscale{1.15}
\includegraphics[width=\columnwidth,trim=2 0 0 -20,clip]{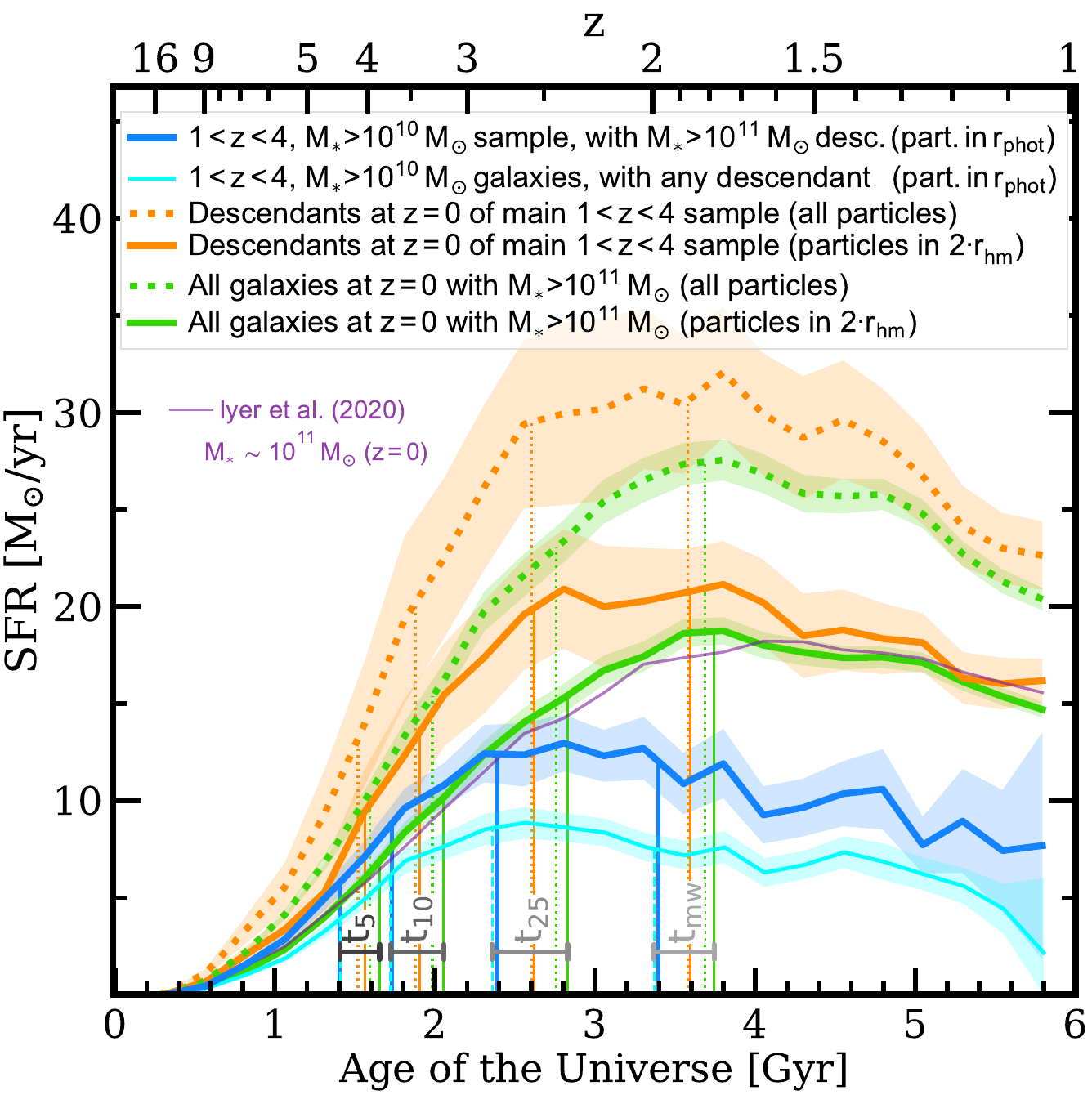}
\caption{Ground-truth median SFHs calculated from the simulated particles for the subsamples of galaxies: all M$_{\ast}>10^{11}$ M$_{\odot}$ galaxies at $z=0$ (green lines), all massive galaxies at $1<z<4$ (cyan), our main $1 < z < 4$ progenitors sample (blue), and the descendants at $z=0$ of this high-redshift sample of progenitors (orange). The results from the 2D-SPS analysis are not included in this figure. The median SFHs for the $z=0$ subsets have been calculated in two ways: considering all the particles in galaxies (dotted lines) or only the particles inside a sphere with radius 2$\times$r$_{\text{hm}}$ to the galaxy center (solid). The median SFH of the two high-redshift samples has been built using only particles whose distance to the galaxy center is lower than the radius of the photometric aperture used for each galaxy. Shaded areas represent the uncertainty of the median. As vertical lines, we show the $t_{\text{5}}$, $t_{\text{10}}$, $t_{\text{25}}$, and $t_{\text{mw}}$ calculated for each SFH shown. As a comparison, we show in purple the median SFH calculated in \citet{2020MNRAS.498..430I} for all galaxies at $z=0$ in the Illustris simulation with M$_{\ast} \sim 10^{11}\,$M$_{\odot}$.\label{fig:res_fig1}}
\end{figure}

Fig.~\ref{fig:res_fig1} shows the sample-averaged (median) SFHs built from the Illustris simulated particles for the four different subsets of galaxies considered: the $1<z<4$ main sample of $1<z<4$ progenitors (blue line), massive galaxies at $1<z<4$ (cyan), the descendants at $z=0$ of our main sample of $1<z<4$ progenitors (orange lines), and all massive galaxies at $z=0$ in Illustris-1 (green lines). To study whether there is any aperture effect on the SFHs, the SFHs of the two subsamples of $z=0$ galaxies have been built by selecting simulated particles at $z=0$ in two different ways: either by considering all the stellar particles belonging to each galaxy (dotted lines), or only the stellar particles inside a sphere with radius 2$\times  r_{\text{hm}}$ (measured at $z=0$; solid lines). In the case of the two high-redshift samples, the radius of each sphere has been set to match the radius of the photometric aperture used for the 2D-SED fits, r$_{\text{phot}}$, and only particles within this radius have been considered. These typical SFHs have been smoothed using a 250$\,$Myr square kernel. Shaded areas correspond to the uncertainty in the median SFH calculated as the 95\% confidence interval: $t_{N-1} \times \sigma /\sqrt{N}$, where $t$ is the $t$ value from the $t$-distribution for 95\% confidence, $\sigma$ has been estimated from the dispersion of the different SFHs, and $N$ is the number of galaxies at each age. We only show median SFHs of the galaxies at $z=0$ down to $z=1$, the lower limit of our main $1<z<4$ sample. Additionally, we point out that the typical SFH of both samples at $1<z<4$ has been calculated from different numbers of galaxies at different ages (what we would be able to perform when working with real data), in contrast to the median SFHs of the other two $z=0$ samples, which count with the same number of galaxies in the entire redshift range.

We remark that no normalization has been applied to the median SFHs shown in Fig.~\ref{fig:res_fig1}. Regarding the different absolute levels of these SFHs, these are a consequence of several factors. First, the difference in SFR values between the dotted lines with respect to the solid lines of the same color, corresponding to the z=0 samples (green and orange) can be explained by the construction method of the SFH followed in each case: SFHs which have been built using all particles present generally higher SFRs than those built using only particles in $2\times$r$_{\text{hm}}$. This is expected, since both for the population of $z=0$ $10^{11}$ M$_{\odot}$ galaxies and for the $z=0$ descendants of the main $1<z<4$ sample, the stellar mass inferred from the typical SFHs of particles within $2\times$r$_{\text{hm}}$ (i.e., the area enclosed by the SFH) is $\sim 66$\% of the stellar mass inferred from the typical SFH built from all the particles in galaxies. We will address the difference between the median SFH of both $z=0$ samples below in this section.  Secondly, the average SFR values from the median SFH calculated for the $z=0$ samples (green and orange) is systematically higher than for the $1 < z < 4$ samples (blue and cyan). This difference is mainly due to the fact that we calculate the SFH for each $z=0$ galaxy from their particles at $z = 0$ instead of computing it after tracing and independently considering the actual precursors of the $z = 0$ galaxy at high-redshift (unfeasible in real observations). As a consequence, the average SFR per $z=0$ galaxy is bigger than the average SFR per precursor, due to the lower number of galaxies considered at $z=0$. In the case of the difference between the median SFH of our main $1 < z < 4$ sample (blue) and their descendants at $z = 0$ (solid orange for particles in $2\times$r$_{\text{hm}}$), there is an additional contribution based on the fact that some of the stellar mass of the $z = 0$ descendants comes from other less massive progenitors which are not included in our main $1 < z < 4$ sample of massive progenitors due to our mass cut-off. In addition to this, the different apertures considered for both $1 < z < 4$ samples with respect to the other $z = 0$ samples also plays a role in lowering the average SFR of these two high-redshift samples: the radii of the photometric apertures used to build the sphere that contains the simulated particles in these high-redshift samples are usually smaller than 2$\times$r$_{\text{hm}}$ measured at the observed redshift, as explained in Section~\ref{subsec:phot}, and should be smaller than 2$\times$r$_{\text{hm}}$ measured at $z=0$. In fact, we remind the reader that the typical photometric aperture radius for our main $1<z<4$ sample of progenitors is nearly 20\% smaller than $2\times $r$_{\text{hm}}$ and encloses around 65\% of the total stellar mass in the galaxies (see Fig.~\ref{fig:rphot_vs_r2hm}). When all massive $1<z<4$ galaxies are considered, these numbers are similar: the photometric aperture has a median radius 19.8\% smaller than $2\times $r$_{\text{hm}}$ and includes $\sim$64.8\% of the stellar mass.

If we integrate the median SFH of both high-redshift samples over cosmic time, the difference in their SFR levels leads to our main $1<z<4$ sample (in blue) recovering a stellar mass of 0.17 dex higher than for all $1<z<4$ massive galaxies (cyan). This is a likely consequence of our main $1<z<4$ sample containing only massive progenitors of local $10^{11}\,$M$_{\odot}$ galaxies, and these progenitors, in order to reach such elevated stellar masses at $z=0$, are expected to be more massive (in average) and have higher SFRs than ordinary massive galaxies at the sample redshift range. In fact, our main $1<z<4$ progenitors sample has a median stellar mass 0.08$\,$dex higher than all $1<z<4$ massive galaxies (for particles inside r$_{\text{phot}}$) and a SFR (in r$_{\text{phot}}$, median and quartiles) of $16_{7}^{31}$ vs$.$ $11_{5}^{23}\,$M$_{\odot}$/yr for all massive galaxies at $1<z<4$.

In Fig.~\ref{fig:res_fig1}, we also show with vertical lines the mass-fraction formation times $t_{\text{5}}$, $t_{\text{10}}$, $t_{\text{25}}$, and $t_{\text{mw}}$ calculated by integrating each sample-averaged SFH over cosmic time. As mentioned before, we only take into account for these calculations the stellar mass formed in each SFH at ages of the Universe below that of the minimum redshift of our main $1<z<4$ sample ($z\sim 1$), as at lower redshifts we lack information on the photometric properties of any of the galaxies in this sample-averaged SFH. This means we only consider the $z>1$ part of the median SFHs for the $z=0$ galaxy subsets and the whole SFH for the high-redshift samples. We find the mass-fraction formation times for the SFH of the high-redshift samples (blue and cyan) are systematically shifted towards earlier times (expressed in terms of the age of the Universe; or correspond to older lookback times), than the ones for the descendants at $z=0$ of our main sample (orange), and these, in turn, are younger than those from the whole population of massive galaxies at $z=0$ (green). These $\sim $200-400$\,$Myr shifts ($\sim10$\% in relative terms) could be explained by a progenitor bias, which appears as a consequence of the mass cutoff of M$_{\ast} > 10^{10}$\,M$_{\odot}$ imposed on our main sample selection. We first focus on the younger (or earlier) formation epochs of the descendants (orange) with respect to those of all the very massive galaxies at $z=0$ (green). The mass cutoff makes the sample of $z=0$ descendants to be biased towards larger masses and more massive galaxies tend to present older stellar population ages. In contrast, when taking into account all M$_{\ast} > 10^{11}$\,M$_{\odot}$ galaxies at $z=0$, i.e., a complete sample of local massive galaxies, we obtain slightly later formation ages. In fact, as it can be seen in Fig.~\ref{fig:res_hists}, the median and quartile stellar masses are $0.1-0.2$ dex larger for the descendants of our sample of M$_{\ast} > 10^{10}$ M$_{\odot}$ galaxies at $1<z<4$ compared to the complete sample of M$_{\ast} > 10^{11}$\,M$_{\odot}$ $z=0$ galaxies. In addition, our mass cut at $1<z<4$ implies losing 17\% of massive M$_{\ast} > 10^{11}$ M$_{\odot}$ galaxies at $z=0$.

Indeed, this progenitor bias also explains why the time-averaged SFR for $z=0$ descendants (orange) is higher than that for the population of very massive galaxies (green). This applies to the galaxy apertures considered in Fig.~\ref{fig:res_fig1} (for which no mass-normalization has been applied). The differences increase for ages corresponding to smaller fractions of the total stellar mass. Furthermore, both for all massive galaxies at $z=0$ and for $z=0$ descendants of the main high-redshift sample, mass-fraction formation times are older (smaller values in age of the Universe) when all stellar particles are included in the SFH computation than when considering only those inside $2\times$r$_{\text{hm}}$. This may be explained by an outward migration of stars, which would result in larger apertures adding a larger fraction of older stars in the SFH computation and, consequently, earlier formation times.  

Regarding the shift to earlier formation times of the main progenitors sample (blue) with respect to the $z=0$ descendants (orange), possibly also due to the progenitor bias, we propose a similar explanation as the one given above. Our main sample of massive progenitors at $1<z<4$ is only a biased subset (cut in mass and with a given redshift distribution) of all the progenitors that evolve to a galaxy from the whole sample of $z=0$ descendants. The descendants, as mentioned above, also have other progenitors that do not fulfill the mass cutoff at $z>1$ and which may have undergone a merger with a galaxy from our main $1<z<4$ sample at a lower redshift. These minor progenitors at $1<z<4$ would not be included in the SFH of the (massive) progenitors sample, and they would probably host younger stellar populations, which would explain the shift towards more recent formation times for their $z=0$ descendants.

Finally, we notice that the ground-truth formation times for our main $1<z<4$ sample of precursors (in blue) are almost identical to the ones derived for all the $1<z<4$ massive galaxies (cyan), with (relative) differences of less than $34\,$Myr (1.4\%) in all cases. This suggests that, although our main progenitors sample accounts for approximately two thirds of all the M$_{\ast} > 10^{10}\,$M$_{\odot}$ galaxies at $1<z<4$, we can estimate the formation times of local M$_{\ast} > 10^{11}$ M$_{\odot}$ descendants by considering all the massive galaxies at the same redshift range.

\newpage
\subsubsection{Recovering the SFH of Massive Galaxies with 2D-SPS Modeling of HST+JWST Data}\label{subsubsec:SFH_2DSPS}

We now calculate the median SFH of the main \mbox{$1<z<4$} galaxy sample obtained from the 2D-SED fits (not from the Illustris database, as done in Section~\ref{subsubsec:particles}). For this calculation, we take the median SFH for each galaxy in this sample constructed from the 2D-SPS analysis (see Section~\ref{subsec:phot_sfhs}), normalize each galaxy SFH to recover the median stellar mass (without remnants or yields) given by its integrated SED-fits, and combine all these normalized galaxy SFHs to build the median (or typical) SFH for the whole sample. We remark that this implies that the SFR at a given age of the Universe involves the combination of a different number of galaxies, namely, all that lie at lower redshifts compared to the redshift corresponding to that age of the Universe. 

\begin{figure}[t]
\centering
\includegraphics[width=\columnwidth,trim=0.2 42 0 0,clip]{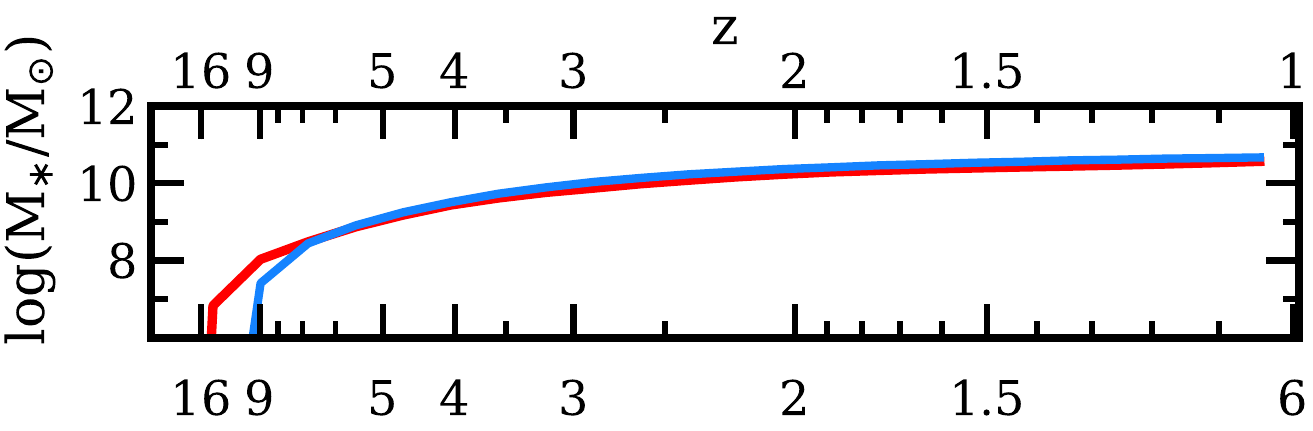}
\includegraphics[width=\columnwidth,trim=0 0 3.4 47,clip]{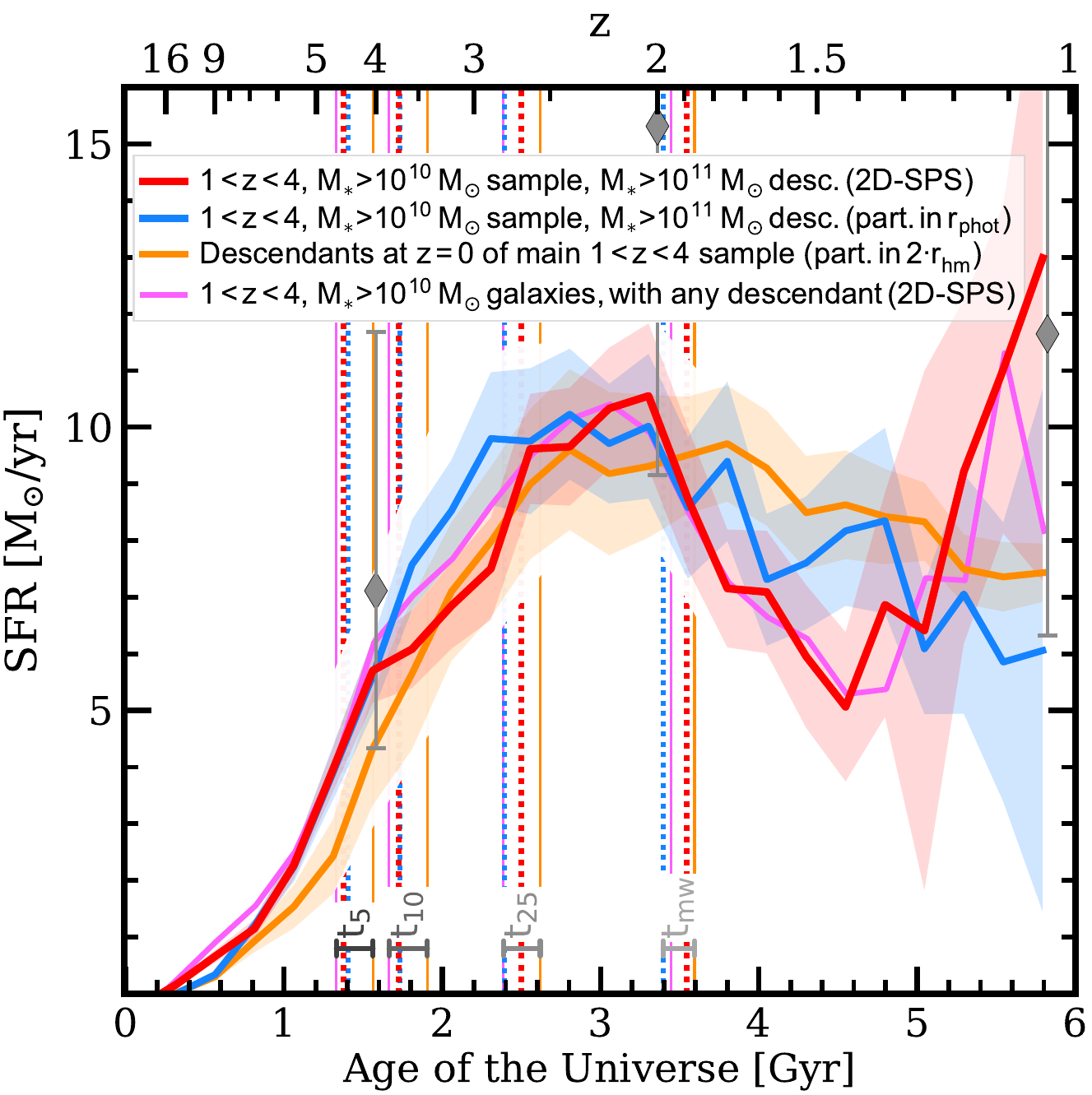}
\caption{Comparison between the median SFH of our main $1<z<4$ progenitors sample derived from the 2D-SED fits (in red) and from the simulated particles (in blue). We also include the median SFH of the descendants of this high-redshift sample at $z=0$ built with particles inside 2·r$_{\text{hm}}$ (in orange) and the median SFH of massive galaxies at $1<z<4$ derived from applying the 2D-SPS method to this sample (pink). Shaded areas represent the uncertainty of the median. The SFHs have been normalized to recover the same stellar mass as the median 2D-SPS-derived SFH of our main $1<z<4$ sample over the same redshift intervals ($z \gtrsim 1$). The vertical lines depict the $t_{\text{5}}$, $t_{\text{10}}$, $t_{\text{25}}$, and $t_{\text{mw}}$ mass-fraction formation times for each SFH. Gray diamonds show the Illustris SFMS level at $z=1$, 2, and 4 (\citealt{2015MNRAS.447.3548S}). On the top panel, we show the evolution of the integrated stellar mass at each redshift for our main high-redshift sample (2D-SPS and particles, same color code).\label{fig:res_fig2}}
\end{figure}

Fig.~\ref{fig:res_fig2} shows this median SFH of the main $1<z<4$ progenitors sample derived from the 2D-SED fits (red solid line). We compare this SFH, which could be obtained following the same procedure on real galaxy samples when JWST (plus HST) data are available, with the ground-truth. First, we compare with the averaged SFH for the same main sample of $1<z<4$ precursors but built from the simulated particles inside the radius of the photometric aperture (blue). This comparison allows to evaluate the accuracy of our method. We also compare with the SFH of $z=0$ descendants using the stellar particles inside 2$\times$r$_{\text{hm}}$ (orange). This comparison allows us to understand the bias of a sample based on imaging data linked to selection effects. The two latter comparison SFHs are the same shown in Fig.~\ref{fig:res_fig1}, but normalized to recover the same stellar mass as the median 2D-SPS-derived SFH of our main $1<z<4$ sample ($10^{10.57}$ M$_{\odot}$ at $z=1$). Note that, without normalization, the corresponding stellar masses at $z=1$ would be $\log($M$_{\ast}$/M$_{\odot})=10.67$ for the median ground-truth SFH of this high-redshift sample and 10.91 for the median SFH of their descendants at $z=0$ (built from particles inside 2$\times$r$_{\text{hm}}$). As commented in Section~\ref{subsubsec:particles}, the difference in the mass recovered by the SFH of our main $1<z<4$ of massive progenitors and the $z=0$ descendants is caused by the reduction in the multiplicity of galaxies when calculating galaxy averages at $z=0$, the presence of less massive galaxy branches for the $z=0$ descendants which are not included in our main $1<z<4$ sample due to the mass cut-off, and the smaller aperture radius (in average) used to calculate the galaxy SFHs for the $1<z<4$ sample. Shaded areas correspond to the uncertainty in the median SFH calculated as the 95\% confidence interval. Finally, we have calculated the median SFH derived from applying the 2D-SPS analysis to all $1<z<4$ massive galaxies in \citetalias{2017MNRAS.468..207S}, following the same procedure than that described for our main $1<z<4$ sample of progenitors. This new median SFH, also normalized to the same mass as the others, is shown in pink. For clarity, we have not included the uncertainties of this last median SFH. 

We find that the general shapes of the four SFHs shown are not very different from each other, except for the SFR increase in the last 0.5-1.0$\,$Gyr (near $z=1$) in the SFHs built from the 2D-SPS analysis (in red and pink). The SFH of our main sample of precursors built from the 2D-SED fits (red) follows the general trend of the SFH built from the simulated particles in the sample (blue), but for this raise, and this also happens for the 2D-SPS-derived SFH of massive galaxies at $1<z<4$ (pink). The number of galaxies considered in the calculation of the SFHs for the high-redshift samples drops as the redshift decreases (as clearly seen in the increase of uncertainties). Thus, there are fewer galaxies near $z=1$ from which to calculate the typical SFH and, when averaging, the weight of any individual galaxy SFH is bigger. The individual galaxy SFHs built from the 2D-SPS, as explained in Section~\ref{subsec:subSPSmodel}, are the combination of the SFH of one young and one old stellar population. This causes the galaxy SFHs built from the 2D-SED fits to normally have a peak in SFR near the redshift of the galaxy, which is not necessary present in the galaxy SFH built from the simulated particles (see, for example, the median 2D-SPS galaxy SFH shown in Fig.~\ref{fig:method}). At higher redshifts, these individual peaks, if present, are not reflected in the median SFHs not only because the SFH is averaged over more galaxies but also because these individual peaks in SFR are located at varying redshifts and distributed across a wide time span. On the contrary, near $z \sim 1$, where the number of galaxies drops, we pile up some of these peaks and this is reflected in the median SFHs from the 2D-SPS analysis.

In Fig.~\ref{fig:res_fig2}, we include the SFR values expected from the Illustris SFMS at $z=1$, 2, and 4 (\citealt{2015MNRAS.447.3548S}) for the stellar masses obtained by integrating our median 2D-SPS SFH of progenitors down to each of those redshifts. We remind the reader that this median 2D-SPS SFH, like other SFHs in Fig.~\ref{fig:res_fig2}, has been normalized to recover the median galaxy stellar mass (without including remnants or yields) of the main $1 < z < 4$ sample of progenitors given by the integrated SED-fits. We find that our median SFR values (in red in Fig.~\ref{fig:res_fig2}) are consistent within the errors with those expected from the SFMS in \citeauthor{2015MNRAS.447.3548S}, although the agreement observed for the $z \sim 1$ SFMS value with respect to our median SFR at that redshift is probably caused by the spurious rise of the SFR in this median SFH at $z \sim 1$ (due to the second peak of star formation in the individual galaxy SFHs). On the top panel, we show the evolution of stellar mass assembly for our main high-redshift sample, calculated both from the median SFH of the 2D-SPS analysis of these progenitors (in red) and that of their stellar particles in r$_{\text{phot}}$ (blue).

The mass-fraction formation times for each SFH in the figure are shown as vertical lines. We remark the good agreement between the $t_{5}$ and $t_{10}$ formation times derived for our main high-redshift sample of precursors using the 2D-SED fits (red) and the simulated particles (blue), with (relative) differences of -29 (-2.1\%) and \mbox{-9\,Myr (-0.5\%)} in age of the Universe, respectively, with the ground-truth values being larger. Additionally, the differences in the other two formation times, although higher, are also low: +107 (+4.5\%) and +149\,Myr (+4.4\%) for $t_{25}$ and $t_{mw}$, respectively, with larger values from the 2D-SED fits. Regarding the comparison of these formation times for the main high-redshift sample (2D-SPS and particles) with their $z=0$ descendants (orange), the agreement is also good, with maximum (relative) differences in age of the Universe of \mbox{-188\,(-12\%)}, \mbox{-184\,(-9.7\%)}, \mbox{-227\,(-8.7\%)}, and \mbox{-199\,Myr\,(-5.5\%)} for $t_{5}$, $t_{10}$, $t_{25}$ and $t_{mw}$, respectively, and \mbox{74\,(5.5\%)}, \mbox{59\,(3.5\%)}, \mbox{112\,(4.7\%)}, and \mbox{98\,Myr\,(2.8\%)} when the main progenitors sample is compared to all massive galaxies at $1<z<4$ (pink). As mentioned before, the older values of the  $t_{5}$, $t_{10}$ and $t_{25}$ formation times for the high-redshift sample with respect to the $z=0$ descendants are due to the progenitor bias. 

According to the 2D-SPS analysis of our main \mbox{$1<z<4$} sample, local M$_{\ast} > 10^{11}$\,M$_{\odot}$ galaxies would have formed 5\% of their stellar mass at $z \sim 1$ by $z=4.5$, 10\% of their mass by $z=3.7$, and 25\% by $z=2.7$. These redshifts correspond to ages of the Universe of 1.4, 1.7, and 2.5\,Gyr, respectively. This is equivalent to saying that 5\% of the stellar mass present in very massive local galaxies at $z \sim 1$ was already assembled when the Universe was only $\sim 10$\% of its current age, 10\% of the mass by 13\% of its age, and 25\% of the mass by 18\% of the cosmic time. 

The first star formation episodes for these galaxies are located at $z=16\pm1$ (calculated as the median redshift where the SFR starts to be larger than 0). Then it would rise up to $z\sim 3$. After that, it would remain approximately constant down to $z\sim2$, and a slightly decreasing trend would follow. This can be clearly seen in the ground-truth SFH derived from the simulated particles in the database and agrees with the median SFHs calculated in \citet{2020MNRAS.498..430I} for \mbox{$10^{10.75} <$ M$_{\ast} < 10^{11.25}$ M$_{\odot}$} galaxies at $z=0$ in the Illustris simulation (considering all the particles in galaxies). Nevertheless, we hardly reproduce the most recent part of the SFH from the 2D-SPS analysis, where our method begins to fail. We note that our 2D-SED fitting method was calibrated in order to successfully reproduce the first instants of the stellar mass assembly in massive galaxies, i.e., $t_{\text{5}}$, $t_{\text{10}}$, $t_{\text{25}}$, along with $t_{\text{mw}}$. Thus, even though we successfully determine these mass-fraction formation times for the typical SFH of massive galaxies, our method fails to reproduce this SFH at higher ages of the Universe. However, the recovery of the whole SFH of massive galaxies and, consequently, their typical SFH throughout the whole redshift range, is beyond the scope of this work. 

\begin{deluxetable*}{cc|rc|cccc|cccc}[ht]
\tablenum{4}
\setlength{\tabcolsep}{2.5pt}
\tablecaption{Mass-fraction formation times and redshifts for different samples \label{tab:tqs}}
\tablehead{\multicolumn{2}{c}{Galaxy subset} & \multicolumn{2}{c}{SFHs from} & \colhead{$t_{\text{5}}$} & \colhead{$t_{\text{10}}$} & \colhead{$t_{\text{25}}$} & \colhead{$t_{\text{mw}}$} & \colhead{$z_{\text{5}}$} & \colhead{$z_{\text{10}}$} & \colhead{$z_{\text{25}}$} & \colhead{$z_{\text{mw}}$} }
\decimals
\startdata
\multirow{2}{*}{(1)} & $1<z<4$, M$_{\ast}>10^{10}\,$M$_{\odot}$ sample & \multicolumn{2}{c|}{2D-SED fits    (r$_{\text{phot}}$)} & \phantom{.S}$1.4_{1.1}^{1.7}$ & \phantom{.S}1.7$_{1.5}^{2.1}$ & \phantom{.S}$2.5_{2.2}^{3.1}$ & \phantom{.S}$3.6_{3.2}^{4.1}$ & \phantom{.S}$4.5_{3.8}^{5.2}$ & \phantom{.S}$3.7_{3.1}^{4.2}$ & \phantom{.S}$2.7_{2.2}^{2.9}$ & \phantom{.S}$1.9_{1.6}^{2.1}$ \\
 & with M$_{\ast}>10^{11}\,$M$_{\odot}$ descendants & Particles & in r$_{\text{phot}}$ & 1.40  & 1.73  & 2.39  & 3.39 & 4.41  & 3.70 & 2.78 & 1.98 \\
\hline
\multirow{2}{*}{(2)} & Descendants at $z=0$ & \multirow{2}{*}{Particles} & all & 1.52 & 1.88 & 2.60 & 3.58 & 4.14 & 3.45 & 2.57 & 1.87 \\
 & of main $1<z<4$ sample    &                            & in 2$\times$r$_{\text{hm}}$ & 1.56 & 1.91 & 2.62 & 3.59 & 4.04 & 3.41 & 2.56 & 1.86 \\
\hline
\multirow{2}{*}{(3)} & All galaxies at $z=0$ & \multirow{2}{*}{Particles} & all & 1.60 & 1.98 & 2.76 & 3.69 & 3.97 & 3.29 & 2.43 & 1.81 \\
 & with M$_{\ast}>10^{11}$ M$_{\odot}$ &  & in 2$\times$r$_{\text{hm}}$ & 1.66 & 2.05 & 2.83 & 3.74 & 3.85 & 3.19 & 2.37 & 1.78 \\
\hline
\multirow{2}{*}{(4)} & $1<z<4$, M$_{\ast}>10^{10}\,$M$_{\odot}$ galaxies & \multicolumn{2}{c|}{2D-SED fits    (r$_{\text{phot}}$)} & 1.33 & 1.66 & 2.39 & 3.45 & 4.61 & 3.83 & 2.79 & 1.95 \\
 & with any descendant & Particles & in r$_{\text{phot}}$ & 1.41 & 1.72  & 2.36  & 3.37 & 4.40  & 3.72 & 2.82 & 1.99
\enddata
\vspace{0.2cm}\scriptsize{\textbf{Note:}\,Values have been calculated from the SFHs shown in Figs$.$~\ref{fig:res_fig1} and~\ref{fig:res_fig2}$.$\,The mass-fraction formation times are measured in Gyr from Big Bang.}
\end{deluxetable*}

\newpage
\vspace{-0.88cm}

We do not find any \textit{starburst epoch} in the typical SFH from the 2D-SPS analysis on the main $1<z<4$ sample of progenitors, and the other SFHs from the simulated particles, albeit 34\% (13\%) of galaxies in our $1<z<4$ progenitors sample reach 50 (100) M$_{\odot}/$yr at some time of their 2D-SPS-derived galaxy SFHs because of the young population assumptions in the SED-fits. These starburst episodes are also present in the ground-truth SFHs of this main $1<z<4$ sample of progenitors with similar numbers: 31\% (14\%) of galaxies reach 50 (100) M$_{\odot}/$yr at some time of their ground-truth galaxy SFH. Starburst events, which are usually short-lived, $\sim$100$\,$Myr (\citealt{2008ApJ...680..246T}, \citealt{2011ApJ...742...96W}, \citealt{2022MNRAS.513.1175E}), have been proven to occur during the evolution of some galaxies, as it can be inferred from the high SFR values of Main Sequence outliers or those of the population of very luminous high-redshift submillimeter galaxies, discovered by \citet{1997ApJ...490L...5S}. The reason why these brief intense star formation episodes are not present in our median SFHs is that they averaged out when we consider a whole population of galaxies, in the same way that they do not appear when calculating the cosmic star formation history \citep{2014ARA&A..52..415M}. In addition to this, the Illustris simulation is known to have has a paucity of strong starburst galaxies, i.e., a fewer fraction of galaxies that lie significantly above the SFMS when compared to observations (\citealt{2015MNRAS.447.3548S}), which appears to be a consequence of the insufficient resolution of Illustris to resolve the sub-kiloparsec starbursting regions (\citealt{2016MNRAS.462.2418S}).

\subsubsection{The\,Variety of\,the Start of SFHs in Massive Galaxies}\label{subsubsec:varietySFHs}

\begin{figure*}[t]
\centering
\includegraphics[width=0.95\textwidth, trim=0 2 0 7,clip]{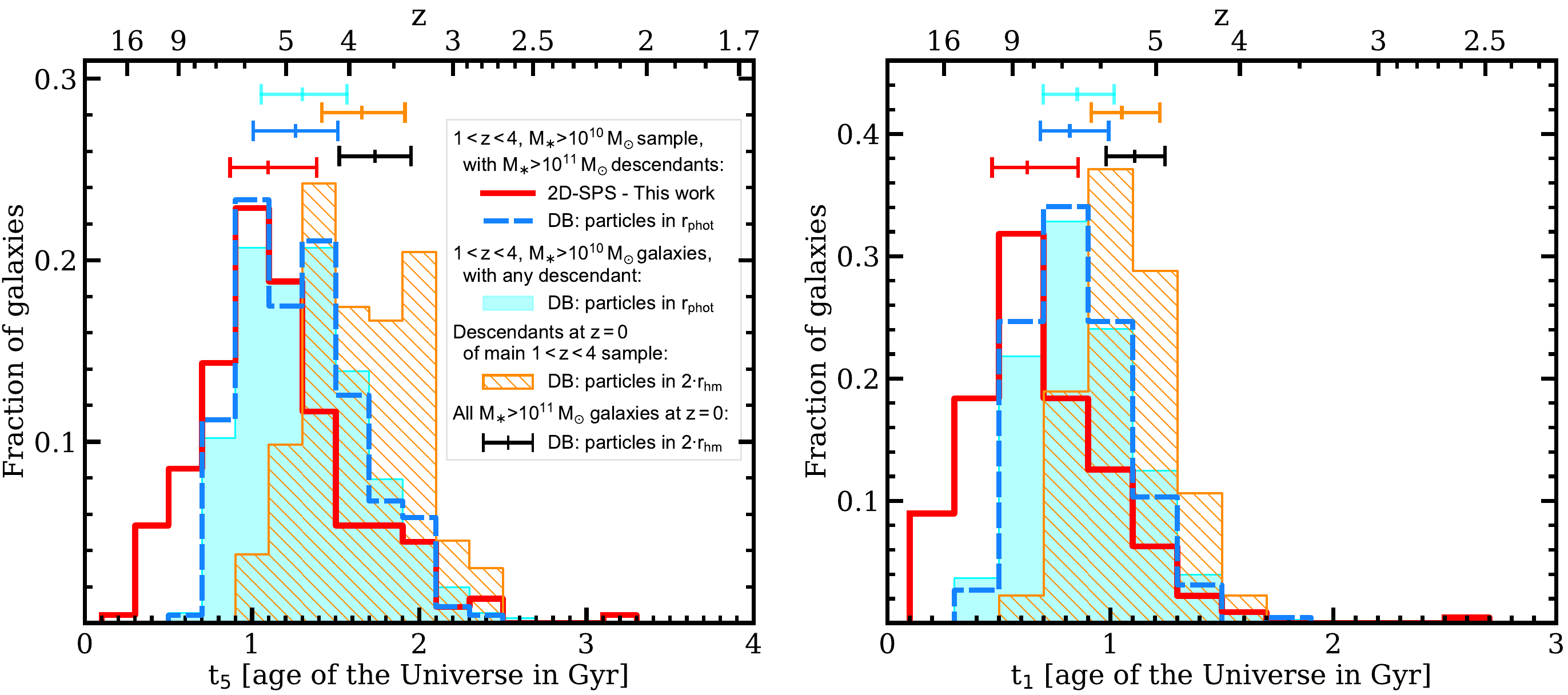}
\caption{Histograms of $t_{5}$ and $t_{1}$ calculated from the SFHs of galaxies in the different samples considered. Our main $1<z<4$ sample of progenitors is represented by the outlined histograms: values calculated from the 2D-SPS-derived galaxy SFHs are outlined in red, while those calculated from the ground-truth SFHs built from the database (DB) are represented with a dashed blue line. For the 2D-SPS-derived values, we only show the median $t_{5}$ or $t_{1}$ for each galaxy (calculated out of the 300 SFHs per galaxy). The cyan, filled histogram represents the distribution of the ground-truth $t_{5}$ and $t_{1}$ for massive galaxies at $1<z<4$, calculated from the ground-truth SFHs built by considering only particles inside r$_{\text{phot}}$ in the database. The ground-truth distribution for the $z=0$ descendants of our main sample is shown as the orange, hatched histogram. For the whole population of very massive galaxies at $z=0$, only its median and quartile values are shown (black segments). For these two $z=0$ samples, only particles inside 2$\times$r$_{\text{hm}}$ have been considered. Median and quartiles are shown as segments on the top.\label{fig:res_t5_t1hist}}
\end{figure*}

Finally, we briefly concentrate on when the star formation started for the individual galaxies in the subsamples. We assume the start of the star formation for each galaxy can be given by its $t_{5}$ (calculated from their individual galaxy SFH) or, even, by its $t_{1}$. 

Fig.~\ref{fig:res_t5_t1hist} shows the diversity in the values of $t_{5}$ (left) and $t_{1}$ (right) for the main sample of progenitors at $1<z<4$ (2D-SPS method and ground-truth from particles in r$_{\text{phot}}$), their descendants at $z=0$ (ground-truth from particles inside $2\times$r$_{hm}$), the whole population of very massive galaxies at $z=0$ (particles inside $2\times$r$_{hm}$), and massive galaxies at $1<z<4$ (ground-truth from particles inside r$_{\text{phot}}$). For clarity, in the case of the population of very massive galaxies at $z=0$ we do not include the histogram but only its statistical values (median and quartiles). The median values of both $t_{5}$ and $t_{1}$ for the main $1<z<4$ progenitors sample given by the 2D-SPS (in red) are consistent with the ground-truth (in blue) for this sample (with differences of 0.26 and 0.19$\,$Gyr, respectively), even though the histogram for our 2D-SPS method predicts that some galaxies start to form a bit earlier ($\sim$0.2-0.4$\,$Gyr earlier) than expected according to the database. These differences are already visible in Fig.~\ref{fig:res_fig2}, where the SFH corresponding to the 2D-SPS analysis (red curve) rises faster in the beginning than that of the simulated particles (in blue). Regarding the distributions of the two subsamples at $z=0$ (descendants of the main $1<z<4$ sample in orange and the whole populations of M$_{\ast}>10^{11}$ M$_{\odot}$ galaxies in black), their median $t_{5}$ and $t_{1}$ are similar and also consistent with each other. This also happens with the ground-truth values of massive galaxies at $1<z<4$ (in cyan) and our main $1<z<4$ sample (in blue), and, consequently, suggests again that the whole population of massive $1<z<4$ galaxies tends to reproduce the earliest formation times of actual massive progenitors (at the same high-redshift range) of M$_{\ast}>10^{11}\,$M$_{\odot}$ galaxies at $z=0$.

Fig.~\ref{fig:res_t5_t1hist} can be useful if we want to know where to look for the first star formation episodes in massive galaxies. According to the ground-truth distribution of $t_{5}$ and $t_{1}$ for the main $1<z<4$ sample of massive progenitors (in blue), 25\% of these galaxies would have formed 1\% (5\%) of its stellar mass by $z \sim 8$ (6), and 50\% of the galaxies by $z \sim 7$ (5). 

In Table~\ref{tab:tqs}, we summarize the mass-fraction formation times (and corresponding redshifts) for all the SFHs shown in Figs.~\ref{fig:res_fig1} and~\ref{fig:res_fig2}. We have calculated an estimation of the uncertainties for the 2D-SPS-derived values of the main sample of $1<z<4$ progenitors by taking into account the (300) individual SFHs for each galaxy and by calculating the mass-fraction formation times for all of them. We estimate the uncertainty for the 2D-SPS-derived formation times as the quartile values of these distributions.

\section{SUMMARY AND CONCLUSIONS}\label{sec:conclusions}

We assess the power of broad-band HST+JWST imaging data for the analysis of the earliest evolutionary phases of galaxies that evolve into local massive objects (M$_{\ast}>10^{11}$ M$_{\odot}$) by analysing the SFHs of progenitors at high redshift in deep cosmological surveys. For that purpose, we apply stellar population synthesis in 2D to a sample of 221 Illustris-1 simulated M$_{\ast}>10^{10}$\,M$_{\odot}$ galaxies at $1<z<4$ that evolve into M$_{\ast}>10^{11}$\,M$_{\odot}$ galaxies at $z=0$. We use ACS, WFC3, and NIRCam data in the optical and near-infrared from the synthetic Illustris-1 deep survey images (\citetalias{2017MNRAS.468..207S}). We measure SEDs for galaxies in the sample from both integrated and 2D multi-wavelength photometry on these images, previously processed to mimic the depths of CANDELS imaging data from HST and ongoing CEERS observations with JWST along with the source extraction and analysis procedures to be applied to data from these surveys. From the SED modeling, we derive the SFH of each source by combining the 2D information from the fits, and compare it with the ground-truth SFH given by the simulated particles belonging to each galaxy in the Illustris-1 database. In this work, we focus on determining the capabilities of broad-band HST+JWST for determining the first episodes in the stellar mass assembly, and our main findings are the following:

(i) We evaluate the success of our 2D-SPS method in recovering the earliest phases of the stellar mass assembly by comparing the mass-fraction formation times $t_{5}$, $t_{10}$, and $t_{25}$ calculated for each galaxy from its 2D-SPS-derived galaxy SFH with those given by its ground-truth SFH. We find that our 2D-SPS method successfully recovers these quantities with a median relative offset between our values and ground-truth of $+4.4$\%, $+1.8$\%, and $-0.1$\% for $t_{5}$, $t_{10}$, and $t_{25}$, respectively, and scatters of 0.21, 0.24, and 0.28 dex. Additionally, no systematic effects are observed as a function of galaxy redshift.

(ii) We build the median SFH of our main \mbox{$1<z<4$} sample of precursors from the 2D-SPS-derived individual galaxy SFHs to infer the mass-fraction formation times of the sample as a whole. Thus, local \mbox{M$_{\ast} > 10^{11}$\,M$_{\odot}$} galaxies would have assembled 5\% of their stellar mass present at $z \sim 1$ by $z=4.5$, 10\% of their mass by $z=3.7$, and 25\% by $z=2.67$, or equivalently, when the age of the Universe was 1.38, 1.72, and 2.50 Gyr, respectively. These ages agree with their ground-truth values derived from the typical SFH built from the simulated particles of galaxies in the sample, with relative differences of -2.1\%, -0.5\%, and +4.5\% for $t_{5}$, $t_{10}$, and $t_{25}$, respectively. Nevertheless, our method fails to reproduce the shape of this SFH for at $z \lesssim 1$, mainly because of the second star formation peaks in the 2D-SPS-derived galaxy SFHs (caused by the second burst assumed in the functional form of the SFHs) which accumulate at this redshift range.

(iii) We compare the formation times derived from the 2D-SPS analysis for this main sample of precursors with the values obtained from the ground-truth median SFH at $z<1$ of local M$_{\ast}>10^{11}$\,M$_{\odot}$ galaxies and of the descendants of this high-redshift sample. The mass-fraction formation epochs from the 2D-SPS are systematically earlier than those inferred from the descendants of this sample. Besides, the latter are higher than the ones inferred from the whole population of M$_{\ast}>10^{11}$ M$_{\odot}$ galaxies. In both cases, these shifts in the formation redshifts, of $\sim$200-400$\,$Myr, are due to the progenitor bias. This bias arises as a consequence of the mass cut-off imposed on the high-redshift progenitor sample, aimed at reproducing the selection suffered in actual HST+JWST survey observations, and which makes this sample to contain only the most massive (and, thus, older) progenitors of local M$_{\ast} > 10^{11}$ M$_{\odot}$ galaxies. Given the way our progenitor sample has been selected, their descendants at $z=0$ also present a median stellar mass that is slightly higher than the whole population of local M$_{\ast} > 10^{11}$ M$_{\odot}$ galaxies, which leads to mass-fraction formation times closer to the Big Bang. 

(iv) With the aim of comparing our results with those that will be inferred from CEERS observations, we perform the same analysis on all the $1 < z < 4$ $10^{10}\,$M$_{\odot}$ galaxies in the processed \citetalias{2017MNRAS.468..207S} images. We find that the formation times derived from the median SFH of this sample are very similar to the ones inferred from our original sample of massive precursors with a local M$_{\ast}>10^{11}\,$M$_{\odot}$ descendant (differences of $<$1.4\% in all the formation times when considering the ground-truth particles and $<$4.7\% from the 2D-SPS analysis). This suggests that we can consider all massive $1 < z < 4$ galaxies observed with CEERS ($+$CANDELS), regardless of their actual $z=0$ descendant, to study the formation times of the most massive descendants at $z = 0$.

(v) Regarding the variety of the $t_{1}$, and $t_{5}$ values, the distribution of these formation times shows that 25\% of our main $1<z<4$ sample of progenitors formed 1\% (5\%) of their stellar mass present at the redshift of observation by $z \sim 8$ (6), and 50\% of the galaxies by $z \sim 7$ (5).

The results from this work show that our 2D-SPS method, when applied to real CANDELS$+$CEERS spatially-resolved broad-band observations of massive $1 < z < 4$ galaxies, will be able to infer when the early stages of the stellar mass assembly took place in these galaxies and, from them, have an estimation of when local $10^{11}\,$M$_{\odot}$ galaxies began to form, within the limitations and biases already discussed throughout this paper. 

We caution the reader that the numerical values of the mass-fraction formation times and other quantities shown in this work are unique to Illustris-1, since they are dependent on the specifications of the simulation, such as the assumed cosmology, the volume and resolution of the simulation, the physical models for galaxy formation (e.g., the star formation and feedback implementations) and on any free parameter. Thus, the values presented in this paper for Illustris-1 are not expected to be necessary similar to those obtained from other simulations like, for example, the IllustrisTNG project (\citealt{2018MNRAS.480.5113M}; \citealt{2018MNRAS.477.1206N}; \citealt{2018MNRAS.475..624N}; \citealt{2018MNRAS.475..648P}; \citealt{2018MNRAS.475..676S}), follow-up of the Illustris simulation which includes as main changes in the physics the incorporation of magnetohydrodynamics and updates in the feedback physics model, among others updates. This new IllustrisTNG series alleviates some of the tensions present between the outcome of the original Illustris simulations and observations (see \citealt{2018MNRAS.473.4077P}), such as, for example, the (too) high cosmic star-formation rate density predicted by Illustris at $z \leq 1$, the excess in the stellar mass function at $z \lesssim 1$ at the low ($\lesssim 10^{10}\,$M$_{\odot}$) and the high ($\gtrsim 10^{11.5}\,$M$_{\odot}$) mass end, the excessively large physical extent for M$_{\ast} \lesssim 10^{10.7}\,$M$_{\odot}$ galaxies (a factor of few larger than observed), or the overpopulation of the blue cloud and green-valley with respect to the red sequence in the galaxy color distribution (see full list in \citealt{2015A&C....13...12N}). An interesting matter to discuss when  HST$+$JWST measurements are available will be to tell whether the prescriptions assumed for Illustris-1, IllustrisTNG or any other simulation are good enough to reproduce the mass-fraction formation times observed for massive high-redshift galaxies or if, on the contrary, they must be used as constraints to refine new galaxy formation models.

\section{Acknowledgments}
We sincerely thank the referee for the thorough reading and insightful comments that helped to improve the quality of this work. This research has been funded by grants PGC2018-093499-B-I00 and RTI2018-096188-B-I00 funded by MCIN/AEI/10.13039/501100011033. \'AGA acknowledges the support of the Universidad Complutense de Madrid through the predoctoral grant CT17/17-CT18/17. AY is supported by an appointment to the NASA Postdoctoral Program (NPP) at NASA Goddard Space Flight Center, administered by Oak Ridge Associated Universities under contract with NASA.  LC acknowledges financial support from Comunidad de Madrid under Atracci\'on de Talento grant 2018-T2/TIC-11612. RMM acknowledge support from Spanish Ministerio de Ciencia, Innovaci\'on y Universidades through grant PGC2018-093499-B-I00, MDM-2017-0737 Unidad de Excelencia ‘‘Maria de Maeztu’’-Centro de Astrobiolog\'ia (INTA-CSIC) by the Spanish Ministry of Science and Innovation/State Agency of Research MCIN/AEI/ 10.13039/501100011033 and by “ERDF A way of making Europe”, and also from the Instituto Nacional de T\'ecnica Aeroespacial SHARDS-JWST project through the PRE-SHARDSJWST/2020 grant.

\vspace{5mm}
\software{synthesizer \citep{2003MNRAS.338..508P,2008ApJ...675..234P}, 
          Photutils \citep{2019zndo...3568287B},
          SExtractor \citep{1996A&AS..117..393B},
          IRAF \citep{1993ASPC...52..173T},
          Astropy \citep{2013A&A...558A..33A, 2018AJ....156..123A}, 
          Colossus \citep{2018ApJS..239...35D},
          NumPy \citep{2011CSE....13b..22V},
          Pandas \citep{2020zndo...3630805R},
          SciPy \citep{2020NatMe..17..261V}, and
          Matplotlib \citep{2007CSE.....9...90H}.}

\bibliographystyle{aasjournal}

\begin{thebibliography}{}
\bibitem[Abdurro'uf et al.(2022)]{2022ApJ...926...81A} Abdurro'uf, Lin, Y.-T., Hirashita, H., et al.\ 2022, \apj, 926, 81. doi:10.3847/1538-4357/ac439a
\bibitem[Akhshik et al.(2022)]{2022arXiv220304979A} Akhshik, M., Whitaker, K.~E., Leja, J., et al.\ 2022, arXiv:2203.04979
\bibitem[Alcalde Pampliega et al.(2019)]{2019ApJ...876..135A} Alcalde Pampliega, B., P{\'e}rez-Gonz{\'a}lez, P.~G., Barro, G., et al.\ 2019, \apj, 876, 135. doi:10.3847/1538-4357/ab14f2
\bibitem[Astropy Collaboration et al.(2013)]{2013A&A...558A..33A} Astropy Collaboration, Robitaille, T.~P., Tollerud, E.~J., et al.\ 2013, \aap, 558, A33. doi:10.1051/0004-6361/201322068
\bibitem[Astropy Collaboration et al.(2018)]{2018AJ....156..123A} Astropy Collaboration, Price-Whelan, A.~M., Sip{\H{o}}cz, B.~M., et al.\ 2018, \aj, 156, 123. doi:10.3847/1538-3881/aabc4f
\bibitem[Beckwith et al.(2006)]{2006AJ....132.1729B} Beckwith, S.~V.~W., Stiavelli, M., Koekemoer, A.~M., et al.\ 2006, \aj, 132, 1729. doi:10.1086/507302
\bibitem[Bellardini et al.(2022)]{2022arXiv220303653B} Bellardini, M.~A., Wetzel, A., Loebman, S.~R., et al.\ 2022, arXiv:2203.03653
\bibitem[Bertin \& Arnouts(1996)]{1996A&AS..117..393B} Bertin, E. \& Arnouts, S.\ 1996, \aaps, 117, 393. doi:10.1051/aas:1996164
\bibitem[Bouwens et al.(2004)]{2004ApJ...611L...1B} Bouwens, R.~J., Illingworth, G.~D., Blakeslee, J.~P., et al.\ 2004, \apjl, 611, L1. doi:10.1086/423786
\bibitem[Bradley et al.(2019)]{2019zndo...3568287B} Bradley, L., Sip{\H{o}}cz, B., Robitaille, T., et al.\ 2019, Zenodo
\bibitem[Bruzual \& Charlot(2003)]{2003MNRAS.344.1000B} Bruzual, G. \& Charlot, S.\ 2003, \mnras, 344, 1000. doi:10.1046/j.1365-8711.2003.06897.x
\bibitem[Calzetti et al.(2000)]{2000ApJ...533..682C} Calzetti, D., Armus, L., Bohlin, R.~C., et al.\ 2000, \apj, 533, 682. doi:10.1086/308692
\bibitem[Chamorro-Cazorla et al.(2022)]{2022A&A...657A..95C} Chamorro-Cazorla, M., Gil de Paz, A., Castillo-Morales, A., et al.\ 2022, \aap, 657, A95. doi:10.1051/0004-6361/202141930
\bibitem[Chabrier(2003)]{2003ApJ...586L.133C} Chabrier, G.\ 2003, \apjl, 586, L133. doi:10.1086/374879
\bibitem[Charlot \& Fall(2000)]{2000ApJ...539..718C} Charlot, S. \& Fall, S.~M.\ 2000, \apj, 539, 718. doi:10.1086/309250
\bibitem[Cook et al.(2016)]{2016ApJ...833..158C} Cook, B.~A., Conroy, C., Pillepich, A., et al.\ 2016, \apj, 833, 158. doi:10.3847/1538-4357/833/2/158
\bibitem[Costantin et al.(2021)]{2021ApJ...913..125C} Costantin, L., P{\'e}rez-Gonz{\'a}lez, P.~G., M{\'e}ndez-Abreu, J., et al.\ 2021, \apj, 913, 125. doi:10.3847/1538-4357/abef72
\bibitem[Costantin et al.(2022)]{2022ApJ...929..121C} Costantin, L., P{\'e}rez-Gonz{\'a}lez, P.~G., M{\'e}ndez-Abreu, J., et al.\ 2022, \apj, 929, 121. doi:10.3847/1538-4357/ac5a57
\bibitem[Cowie et al.(1996)]{1996AJ....112..839C} Cowie, L.~L., Songaila, A., Hu, E.~M., et al.\ 1996, \aj, 112, 839. doi:10.1086/118058
\bibitem[Davison et al.(2020)]{2020MNRAS.497...81D} Davison, T.~A., Norris, M.~A., Pfeffer, J.~L., et al.\ 2020, \mnras, 497, 81. doi:10.1093/mnras/staa1816
\bibitem[Dekel \& Burkert(2014)]{2014MNRAS.438.1870D} Dekel, A. \& Burkert, A.\ 2014, \mnras, 438, 1870. doi:10.1093/mnras/stt2331
\bibitem[De Lucia \& Blaizot(2007)]{2007MNRAS.375....2D} De Lucia, G. \& Blaizot, J.\ 2007, \mnras, 375, 2. doi:10.1111/j.1365-2966.2006.11287.x
\bibitem[Diemer(2018)]{2018ApJS..239...35D} Diemer, B.\ 2018, \apjs, 239, 35. doi:10.3847/1538-4365/aaee8c
\bibitem[Di Matteo et al.(2013)]{2013A&A...553A.102D} Di Matteo, P., Haywood, M., Combes, F., et al.\ 2013, \aap, 553, A102. doi:10.1051/0004-6361/201220539
\bibitem[Dom{\'\i}nguez S{\'a}nchez et al.(2016)]{2016MNRAS.457.3743D} Dom{\'\i}nguez S{\'a}nchez, H., P{\'e}rez-Gonz{\'a}lez, P.~G., Esquej, P., et al.\ 2016, \mnras, 457, 3743. doi:10.1093/mnras/stw201
\bibitem[Dubois et al.(2016)]{2016MNRAS.463.3948D} Dubois, Y., Peirani, S., Pichon, C., et al.\ 2016, \mnras, 463, 3948. doi:10.1093/mnras/stw2265
\bibitem[Dunlop et al.(2021)]{2021jwst.prop.1837D} Dunlop, J.~S., Abraham, R.~G., Ashby, M.~L.~N., et al.\ 2021, JWST Proposal. Cycle 1, 1837
\bibitem[Elias et al.(2018)]{2018MNRAS.479.4004E} Elias, L.~M., Sales, L.~V., Creasey, P., et al.\ 2018, \mnras, 479, 4004. doi:10.1093/mnras/sty1718
\bibitem[Ellis et al.(2013)]{2013ApJ...763L...7E} Ellis, R.~S., McLure, R.~J., Dunlop, J.~S., et al.\ 2013, \apjl, 763, L7. doi:10.1088/2041-8205/763/1/L7
\bibitem[Erb et al.(2006)]{2006ApJ...644..813E} Erb, D.~K., Shapley, A.~E., Pettini, M., et al.\ 2006, \apj, 644, 813. doi:10.1086/503623
\bibitem[Espino-Briones et al.(2022)]{2022MNRAS.513.1175E} Espino-Briones, N., P{\'e}rez-Gonz{\'a}lez, P.~G., Zamorano, J., et al.\ 2022, \mnras, 513, 1175. doi:10.1093/mnras/stac728
\bibitem[Finkelstein et al.(2017)]{2017jwst.prop.1345F} Finkelstein, S.~L., Dickinson, M., Ferguson, H.~C., et al.\ 2017, JWST Proposal ID 1345. Cycle 0 Early Release Science, 1345
\bibitem[Forrest et al.(2020a)]{2020ApJ...890L...1F} Forrest, B., Annunziatella, M., Wilson, G., et al.\ 2020, \apjl, 890, L1. doi:10.3847/2041-8213/ab5b9f
\bibitem[Forrest et al.(2020b)]{2020ApJ...903...47F} Forrest, B., Marsan, Z.~C., Annunziatella, M., et al.\ 2020, \apj, 903, 47. doi:10.3847/1538-4357/abb819
\bibitem[Gardner et al.(2006)]{2006SSRv..123..485G} Gardner, J.~P., Mather, J.~C., Clampin, M., et al.\ 2006, \ssr, 123, 485. doi:10.1007/s11214-006-8315-7
\bibitem[Gil de Paz \& Madore(2002)]{2002AJ....123.1864G} Gil de Paz, A. \& Madore, B.~F.\ 2002, \aj, 123, 1864. doi:10.1086/339480
\bibitem[Glazebrook et al.(2017)]{2017Natur.544...71G} Glazebrook, K., Schreiber, C., Labb{\'e}, I., et al.\ 2017, \nat, 544, 71. doi:10.1038/nature21680
\bibitem[Grogin et al.(2011)]{2011ApJS..197...35G} Grogin, N.~A., Kocevski, D.~D., Faber, S.~M., et al.\ 2011, \apjs, 197, 35. doi:10.1088/0067-0049/197/2/35
\bibitem[Ho et al.(2018)]{2018A&A...618A..64H} Ho, I.-T., Meidt, S.~E., Kudritzki, R.-P., et al.\ 2018, \aap, 618, A64. doi:10.1051/0004-6361/201833262
\bibitem[Hunter(2007)]{2007CSE.....9...90H} Hunter, J.~D.\ 2007, Computing in Science and Engineering, 9, 90. doi:10.1109/MCSE.2007.55
\bibitem[Illingworth et al.(2013)]{2013ApJS..209....6I} Illingworth, G.~D., Magee, D., Oesch, P.~A., et al.\ 2013, \apjs, 209, 6. doi:10.1088/0067-0049/209/1/6
\bibitem[Iyer et al.(2020)]{2020MNRAS.498..430I} Iyer, K.~G., Tacchella, S., Genel, S., et al.\ 2020, \mnras, 498, 430. doi:10.1093/mnras/staa2150
\bibitem[Ji \& Giavalisco(2022)]{2022arXiv220402414J} Ji, Z. \& Giavalisco, M.\ 2022, arXiv:2204.02414
\bibitem[Johnston et al.(2022)]{2022MNRAS.tmp.1686J} Johnston, E.~J., H{\"a}u{\ss}ler, B., \& Jegatheesan, K.\ 2022, \mnras. doi:10.1093/mnras/stac1725
\bibitem[Jonsson(2006)]{2006MNRAS.372....2J} Jonsson, P.\ 2006, \mnras, 372, 2. doi:10.1111/j.1365-2966.2006.10884.x
\bibitem[Jonsson et al.(2010)]{2010MNRAS.403...17J} Jonsson, P., Groves, B.~A., \& Cox, T.~J.\ 2010, \mnras, 403, 17. doi:10.1111/j.1365-2966.2009.16087.x
\bibitem[Kitzbichler \& White(2007)]{2007MNRAS.376....2K} Kitzbichler, M.~G. \& White, S.~D.~M.\ 2007, \mnras, 376, 2. doi:10.1111/j.1365-2966.2007.11458.x
\bibitem[Koekemoer et al.(2011)]{2011ApJS..197...36K} Koekemoer, A.~M., Faber, S.~M., Ferguson, H.~C., et al.\ 2011, \apjs, 197, 36. doi:10.1088/0067-0049/197/2/36
\bibitem[Koekemoer et al.(2013)]{2013ApJS..209....3K} Koekemoer, A.~M., Ellis, R.~S., McLure, R.~J., et al.\ 2013, \apjs, 209, 3. doi:10.1088/0067-0049/209/1/3
\bibitem[Kroupa(2001)]{2001MNRAS.322..231K} Kroupa, P.\ 2001, \mnras, 322, 231. doi:10.1046/j.1365-8711.2001.04022.x
\bibitem[Lee \& Yi(2013)]{2013ApJ...766...38L} Lee, J. \& Yi, S.~K.\ 2013, \apj, 766, 38. doi:10.1088/0004-637X/766/1/38
\bibitem[Lee \& Yi(2017)]{2017ApJ...836..161L} Lee, J. \& Yi, S.~K.\ 2017, \apj, 836, 161. doi:10.3847/1538-4357/aa5b87
\bibitem[Leitherer et al.(1999)]{1999ApJS..123....3L} Leitherer, C., Schaerer, D., Goldader, J.~D., et al.\ 1999, \apjs, 123, 3. doi:10.1086/313233
\bibitem[Madau \& Dickinson(2014)]{2014ARA&A..52..415M} Madau, P. \& Dickinson, M.\ 2014, \araa, 52, 415. doi:10.1146/annurev-astro-081811-125615
\bibitem[Maiolino et al.(2008)]{2008A&A...488..463M} Maiolino, R., Nagao, T., Grazian, A., et al.\ 2008, \aap, 488, 463. doi:10.1051/0004-6361:200809678
\bibitem[Mannucci et al.(2009)]{2009MNRAS.398.1915M} Mannucci, F., Cresci, G., Maiolino, R., et al.\ 2009, \mnras, 398, 1915. doi:10.1111/j.1365-2966.2009.15185.x
\bibitem[Marinacci et al.(2018)]{2018MNRAS.480.5113M} Marinacci, F., Vogelsberger, M., Pakmor, R., et al.\ 2018, \mnras, 480, 5113. doi:10.1093/mnras/sty2206
\bibitem[Marsan et al.(2022)]{2022ApJ...924...25M} Marsan, Z.~C., Muzzin, A., Marchesini, D., et al.\ 2022, \apj, 924, 25. doi:10.3847/1538-4357/ac312a
\bibitem[M{\'e}ndez-Abreu et al.(2021)]{2021MNRAS.504.3058M} M{\'e}ndez-Abreu, J., de Lorenzo-C{\'a}ceres, A., \& S{\'a}nchez, S.~F.\ 2021, \mnras, 504, 3058. doi:10.1093/mnras/stab1064
\bibitem[Naiman et al.(2018)]{2018MNRAS.477.1206N} Naiman, J.~P., Pillepich, A., Springel, V., et al.\ 2018, \mnras, 477, 1206. doi:10.1093/mnras/sty618
\bibitem[Nelson et al.(2015)]{2015A&C....13...12N} Nelson, D., Pillepich, A., Genel, S., et al.\ 2015, Astronomy and Computing, 13, 12. doi:10.1016/j.ascom.2015.09.003
\bibitem[Nelson et al.(2016)]{2016ApJ...828...27N} Nelson, E.~J., van Dokkum, P.~G., F{\"o}rster Schreiber, N.~M., et al.\ 2016, \apj, 828, 27. doi:10.3847/0004-637X/828/1/27
\bibitem[Nelson et al.(2018)]{2018MNRAS.475..624N} Nelson, D., Pillepich, A., Springel, V., et al.\ 2018, \mnras, 475, 624. doi:10.1093/mnras/stx3040
\bibitem[Norgaard-Nielsen \& Perez-Gonzalez(2017)]{2017jwst.prop.1283N} Norgaard-Nielsen, H.~U. \& Perez-Gonzalez, P.~G.\ 2017, JWST Proposal. Cycle 1, 1283
\bibitem[Oesch et al.(2010a)]{2010ApJ...709L..16O} Oesch, P.~A., Bouwens, R.~J., Illingworth, G.~D., et al.\ 2010, \apjl, 709, L16. doi:10.1088/2041-8205/709/1/L16
\bibitem[Oesch et al.(2010b)]{2010ApJ...709L..21O} Oesch, P.~A., Bouwens, R.~J., Carollo, C.~M., et al.\ 2010, \apjl, 709, L21. doi:10.1088/2041-8205/709/1/L21
\bibitem[Oke \& Gunn(1983)]{1983ApJ...266..713O} Oke, J.~B. \& Gunn, J.~E.\ 1983, \apj, 266, 713. doi:10.1086/160817
\bibitem[Ono et al.(2013)]{2013ApJ...777..155O} Ono, Y., Ouchi, M., Curtis-Lake, E., et al.\ 2013, \apj, 777, 155. doi:10.1088/0004-637X/777/2/155
\bibitem[Oser et al.(2010)]{2010ApJ...725.2312O} Oser, L., Ostriker, J.~P., Naab, T., et al.\ 2010, \apj, 725, 2312. doi:10.1088/0004-637X/725/2/2312
\bibitem[P{\'e}rez-Gonz{\'a}lez et al.(2003)]{2003MNRAS.338..508P} P{\'e}rez-Gonz{\'a}lez, P.~G., Gil de Paz, A., Zamorano, J., et al.\ 2003, \mnras, 338, 508. doi:10.1046/j.1365-8711.2003.06077.x
\bibitem[P{\'e}rez-Gonz{\'a}lez et al.(2008)]{2008ApJ...675..234P} P{\'e}rez-Gonz{\'a}lez, P.~G., Rieke, G.~H., Villar, V., et al.\ 2008, \apj, 675, 234. doi:10.1086/523690
\bibitem[Pillepich et al.(2018a)]{2018MNRAS.473.4077P} Pillepich, A., Springel, V., Nelson, D., et al.\ 2018, \mnras, 473, 4077. doi:10.1093/mnras/stx2656
\bibitem[Pillepich et al.(2018b)]{2018MNRAS.475..648P} Pillepich, A., Nelson, D., Hernquist, L., et al.\ 2018, \mnras, 475, 648. doi:10.1093/mnras/stx3112
\bibitem[Pontoppidan et al.(2016)]{2016SPIE.9910E..16P} Pontoppidan, K.~M., Pickering, T.~E., Laidler, V.~G., et al.\ 2016, \procspie, 9910, 991016. doi:10.1117/12.2231768
\bibitem[Pulsoni et al.(2021)]{2021A&A...647A..95P} Pulsoni, C., Gerhard, O., Arnaboldi, M., et al.\ 2021, \aap, 647, A95. doi:10.1051/0004-6361/202039166
\bibitem[Reback et al.(2020)]{2020zndo...3630805R} Reback, J., McKinney, W., Jbrockmendel, et al.\ 2020, Zenodo
\bibitem[Remus \& Forbes(2022)]{2022ApJ...935...37R} Remus, R.-S. \& Forbes, D.~A.\ 2022, \apj, 935, 37. doi:10.3847/1538-4357/ac7b30
\bibitem[Renzini(2006)]{2006ARA&A..44..141R} Renzini, A.\ 2006, \araa, 44, 141. doi:10.1146/annurev.astro.44.051905.092450
\bibitem[Rieke et al.(2019)]{2019BAAS...51c..45R} Rieke, M., Arribas, S., Bunker, A., et al.\ 2019, \baas, 51, 45
\bibitem[Rodriguez-Gomez et al.(2015)]{2015MNRAS.449...49R} Rodriguez-Gomez, V., Genel, S., Vogelsberger, M., et al.\ 2015, \mnras, 449, 49. doi:10.1093/mnras/stv264
\bibitem[Rodriguez-Gomez et al.(2016)]{2016MNRAS.458.2371R} Rodriguez-Gomez, V., Pillepich, A., Sales, L.~V., et al.\ 2016, \mnras, 458, 2371. doi:10.1093/mnras/stw456
\bibitem[Sales et al.(2015)]{2015MNRAS.447L...6S} Sales, L.~V., Vogelsberger, M., Genel, S., et al.\ 2015, \mnras, 447, L6. doi:10.1093/mnrasl/slu173
\bibitem[S{\'a}nchez-Menguiano et al.(2020)]{2020MNRAS.492.4149S} S{\'a}nchez-Menguiano, L., S{\'a}nchez, S.~F., P{\'e}rez, I., et al.\ 2020, \mnras, 492, 4149. doi:10.1093/mnras/staa088
\bibitem[Skelton et al.(2014)]{2014ApJS..214...24S} Skelton, R.~E., Whitaker, K.~E., Momcheva, I.~G., et al.\ 2014, \apjs, 214, 24. doi:10.1088/0067-0049/214/2/24
\bibitem[Smail et al.(1997)]{1997ApJ...490L...5S} Smail, I., Ivison, R.~J., \& Blain, A.~W.\ 1997, \apjl, 490, L5. doi:10.1086/311017
\bibitem[Snyder et al.(2015)]{2015MNRAS.454.1886S} Snyder, G.~F., Torrey, P., Lotz, J.~M., et al.\ 2015, \mnras, 454, 1886. doi:10.1093/mnras/stv2078
\bibitem[Snyder et al.(2017)]{2017MNRAS.468..207S} Snyder, G.~F., Lotz, J.~M., Rodriguez-Gomez, V., et al.\ 2017, \mnras, 468, 207. doi:10.1093/mnras/stx487
\bibitem[Sparre et al.(2015)]{2015MNRAS.447.3548S} Sparre, M., Hayward, C.~C., Springel, V., et al.\ 2015, \mnras, 447, 3548. doi:10.1093/mnras/stu2713
\bibitem[Sparre \& Springel(2016)]{2016MNRAS.462.2418S} Sparre, M. \& Springel, V.\ 2016, \mnras, 462, 2418. doi:10.1093/mnras/stw1793
\bibitem[Springel et al.(2018)]{2018MNRAS.475..676S} Springel, V., Pakmor, R., Pillepich, A., et al.\ 2018, \mnras, 475, 676. doi:10.1093/mnras/stx3304
\bibitem[Tacchella et al.(2015)]{2015Sci...348..314T} Tacchella, S., Carollo, C.~M., Renzini, A., et al.\ 2015, Science, 348, 314. doi:10.1126/science.1261094
\bibitem[Tacchella et al.(2016)]{2016MNRAS.458..242T} Tacchella, S., Dekel, A., Carollo, C.~M., et al.\ 2016, \mnras, 458, 242. doi:10.1093/mnras/stw303
\bibitem[Tacchella et al.(2018)]{2018ApJ...859...56T} Tacchella, S., Carollo, C.~M., F{\"o}rster Schreiber, N.~M., et al.\ 2018, \apj, 859, 56. doi:10.3847/1538-4357/aabf8b
\bibitem[Tacchella et al.(2019)]{2019MNRAS.487.5416T} Tacchella, S., Diemer, B., Hernquist, L., et al.\ 2019, \mnras, 487, 5416. doi:10.1093/mnras/stz1657
\bibitem[Tacconi et al.(2008)]{2008ApJ...680..246T} Tacconi, L.~J., Genzel, R., Smail, I., et al.\ 2008, \apj, 680, 246. doi:10.1086/587168
\bibitem[Tody(1993)]{1993ASPC...52..173T} Tody, D.\ 1993, Astronomical Data Analysis Software and Systems II, 52, 173
\bibitem[Torrey et al.(2015)]{2015MNRAS.447.2753T} Torrey, P., Snyder, G.~F., Vogelsberger, M., et al.\ 2015, \mnras, 447, 2753. doi:10.1093/mnras/stu2592
\bibitem[Valentino et al.(2020)]{2020ApJ...889...93V} Valentino, F., Tanaka, M., Davidzon, I., et al.\ 2020, \apj, 889, 93. doi:10.3847/1538-4357/ab64dc
\bibitem[van der Walt et al.(2011)]{2011CSE....13b..22V} van der Walt, S., Colbert, S.~C., \& Varoquaux, G.\ 2011, Computing in Science and Engineering, 13, 22. doi:10.1109/MCSE.2011.37
\bibitem[Virtanen et al.(2020)]{2020NatMe..17..261V} Virtanen, P., Gommers, R., Oliphant, T.~E., et al.\ 2020, Nature Methods, 17, 261. doi:10.1038/s41592-019-0686-2
\bibitem[Vogelsberger et al.(2014a)]{2014Natur.509..177V} Vogelsberger, M., Genel, S., Springel, V., et al.\ 2014, \nat, 509, 177. doi:10.1038/nature13316
\bibitem[Vogelsberger et al.(2014b)]{2014MNRAS.444.1518V} Vogelsberger, M., Genel, S., Springel, V., et al.\ 2014, \mnras, 444, 1518. doi:10.1093/mnras/stu1536
\bibitem[Wang et al.(2017)]{2017MNRAS.469.4063W} Wang, W., Faber, S.~M., Liu, F.~S., et al.\ 2017, \mnras, 469, 4063. doi:10.1093/mnras/stx1148
\bibitem[Williams et al.(2018)]{2018ApJS..236...33W} Williams, C.~C., Curtis-Lake, E., Hainline, K.~N., et al.\ 2018, \apjs, 236, 33. doi:10.3847/1538-4365/aabcbb
\bibitem[Wuyts et al.(2011)]{2011ApJ...742...96W} Wuyts, S., F{\"o}rster Schreiber, N.~M., van der Wel, A., et al.\ 2011, \apj, 742, 96. doi:10.1088/0004-637X/742/2/96
\bibitem[Zahid et al.(2011)]{2011ApJ...730..137Z} Zahid, H.~J., Kewley, L.~J., \& Bresolin, F.\ 2011, \apj, 730, 137. doi:10.1088/0004-637X/730/2/137
\bibitem[Zolotov et al.(2015)]{2015MNRAS.450.2327Z} Zolotov, A., Dekel, A., Mandelker, N., et al.\ 2015, \mnras, 450, 2327. doi:10.1093/mnras/stv740
\end{thebibliography}

\end{document}